# On Operator Mixing in $\mathcal{N} = 4$ SYM

Massimo Bianchi, Burkhard Eden, Giancarlo Rossi and Yassen S. Stanev[†]

*Dipartimento di Fisica, Università di Roma "Tor Vergata"*
*I.N.F.N. – Sezione di Roma "Tor Vergata"*
*Via della Ricerca Scientifica, 1*
*00133 Roma, ITALY*

**Abstract**

We resolve the mixing of the scalar operators of naive dimension 4 belonging to the representation $\mathbf{20'}$ of the $SU(4)$ R–symmetry in $\mathcal{N} = 4$ SYM. We compute the order $g^2$ corrections to their anomalous dimensions and show the absence of instantonic contributions thereof. Ratios of the resulting expressions are irrational numbers, even in the large $N$ limit where, however, we observe the expected decoupling of double-trace operators from single-trace ones. We briefly comment on the generalizations of our results required in order to make contact with the double scaling limit of the theory conjectured to be holographically dual to type IIB superstring on a pp-wave.

---

[†] On leave of absence from Institute for Nuclear Research and Nuclear Energy, Bulgarian Academy of Sciences, BG-1784, Sofia, Bulgaria

# 1 Introduction and summary

The holographic correspondence between superstring theory in anti de Sitter (AdS) spaces and superconformal theories (CFT) [1, 2, 3] has renewed the interest in $\mathcal{N}=4$ supersymmetric Yang-Mills theory (SYM) and triggered the discovery of new unexpected properties of its superconformal phase.

Until very recently, most of the available results on the superstring side were confined to the low energy (weak curvature $L^2 >> \alpha'$) approximation where supergravity takes over. This limit corresponds to the strong 't Hooft coupling ($g^2 N >> 1$) regime of the dual gauge theory which is obviously inaccessible by perturbative means. Barring few important exceptions, perturbative and non-perturbative tests [4] were thus restricted to protected quantities, *i.e.* observables which are actually independent of the coupling constant. By now there is quite a long list of such protected quantities that includes dimensions of operators belonging to short multiplets [5], certain OPE coefficients of chiral primary operators (CPO's) [6], extremal [7] and next-to-extremal [8] correlators.

The first truly dynamical test of the correspondence emerged from the remarkable agreement between SYM instanton effects and D-instanton corrections to higher derivative terms in the type IIB superstring effective action [9]. Bonus symmetry [10] of up to four-point functions of protected operators was another suggestive hint to the underlying type IIB string description of $\mathcal{N}=4$ SYM. Another class of observables that should clearly display stringy behaviour are Maldacena-Wilson loops [11]. So far non-trivial string predictions [12] have only received partial support from perturbation theory [13, 14] and seem to require a deeper understanding of D-instanton effects in order to accommodate SYM instanton corrections [15]. Among the other achievements of $\mathcal{N}=4$ super-instanton calculus the 'two-line proof' [4] of the partial non-renormalization of the four-point function of operators in the supercurrent multiplet [16] stands out for its simplicity. However, partial non-renormalization is essentially a consequence of $SU(2,2|4)$ superconformal symmetry that severely constrains the dynamics, though it does not completely trivialize it [16, 17].

The main purpose of this paper is to resolve the mixing of the scalar primary operators of naive dimension $\Delta_0 = 4$ in the (real) representation $\mathbf{20'}$ of the $SU(4)$ R-symmetry group and compute their anomalous dimensions at order $g^2$.

Relying on previous results on four-point functions of lowest CPO's $\mathcal{Q}$ with [18] or without [19, 20, 21] insertion of the lowest Konishi scalar $\mathcal{K}$, we disentangle the mixing among the scalar operators in the $\mathbf{20'}$ representation. The vanishing of instanton contributions to the relevant four-point functions [18] implies the absence of non-perturbative corrections to the mixing coefficients and anomalous dimensions that we compute. Our analysis is further simplified by the observation that operators which belong to the Konishi multiplet and whose leading terms are generalized Yukawa couplings, decouple at the order at which we work. The remaining operators show an intricate pattern of mixing at finite $N$ that simplifies significantly as $N \to \infty$. In this analysis we exploit the vanishing of the anomalous dimension of the operator $\mathcal{D}_{\mathbf{20'}}$ that appears in OPE of two $\mathcal{Q}$'s. This property may be viewed as resulting from a generalized shortening condition of the 'linear' type that survives interaction as a consequence of certain differential constraints satisfied by three point functions involving two protected operators [22, 23] [1]. Our results for the one-loop anomalous dimensions definitely exclude the possibility [26] that ratios of anomalous dimensions be rational even in the large $N$ limit. This is not in conflict with any basic principle but rather suggests that the theory behaves in a highly non trivial fashion. Still, it is reassuring to find that multi-trace operators that are dual to multi-particle bound states have in the limit $N \to \infty$ with $g^2 N$ fixed anomalous dimensions that are given by the sums of the anomalous dimensions of their constituents. This suggests that the dual bound states are at threshold. Moreover, as observed in [18], the absence of non-perturbative instanton corrections for the anomalous dimensions and OPE coefficients of the operators which we study is in line with S-duality that maps operators dual to string excitations into operators dual to dyonic string excitations.

The pattern of intricate mixings and irrational anomalous dimensions that we find, may not necessarily prove to be an unsurmountable obstacle towards the extrapolation of the string spectrum and interactions from low energy (strong coupling) to large curvature (weak coupling) at least in the Penrose limit of $AdS_5 \times S^5$ which gives rise to a maximally supersymmetric pp-wave [27, 28]. String loop corrections, which seem to be calculable in the pp-wave background [29], may play a crucial role in quantitatively establishing this correspondence [30, 31]. We will argue that our results can be generalized to yield further insight into the properties of the set of operators dual to the low-lying string excitations.

The plan of the paper is as follows: After recalling some basic definitions and establishing our notation in Section 2, in Section 3 we briefly describe unitary irreducible representations (UIR's) of the superconformal group $SU(2,2|4)$ and discuss the emergence of multiplet shortening. In Section 4 we identify the scalar composite operators of naive dimension $\Delta_0 = 4$ belonging to the representation $\mathbf{20}'$. In Section 5 we perform to order $g^2$ the orthogonalization of the two-point functions and compute the anomalous dimensions of these operators. In Sections 6 and 7 we follow a different route to the same results that requires the computation of the four-point function of two $\mathcal{Q}$'s and two $\mathcal{K}$'s at order $g^4$, thus extending similar results previously obtained at order $g^2$ [18]. In Section 8 we briefly comment on possible generalizations of our results to the double scaling limit of the theory which is conjectured to be holographically dual to type IIB superstring on a pp-wave [27]. In the Appendix we gather unwieldy formulae.

## 2 Notation and conventions

In this section, we summarize our notations and conventions and we recall some relevant results of [18] concerning the structure and the renormalization properties of the unprotected Konishi supermultiplet.

The field content of $\mathcal{N} = 4$ SYM [32] comprises a vector, $A_\mu$, four Weyl spinors, $\psi^A$ ($A = 1, 2, 3, 4$), and six real scalars, $\varphi^i$ ($i = 1, 2, \ldots, 6$), all in the adjoint representation of the gauge group, that we take to be $SU(N)$ for definiteness. In the $\mathcal{N} = 1$ approach that we shall follow the fundamental fields can be arranged into a vector superfield, $V$, and three chiral superfields, $\Phi^I$ ($I = 1, 2, 3$). The six real scalars, $\varphi^i$, are combined into three complex fields, $\phi^I = (\varphi^I + i\varphi^{I+3})/\sqrt{2}$ and $\phi_I^\dagger = (\varphi^I - i\varphi^{I+3})/\sqrt{2}$ that are the lowest components of the chiral and antichiral superfields, $\Phi^I$ and $\Phi_I^\dagger$, respectively. Three of the Weyl fermions, $\psi^I$, are the spinors of the chiral multiplets. The fourth spinor, $\lambda = \psi^4$, together with the vector, $A_\mu$, form the vector multiplet. In this way only an $SU(3) \otimes U(1)$ subgroup of the full $SU(4)$ R-symmetry is manifest.

---

[1]It should be kept in mind, however, that there are operators that satisfy the same shortening condition, *i.e.* saturate the same unitarity bound at tree level, but violate it after inclusion of radiative corrections [24, 25].

The complete $\mathcal{N} = 4$ SYM action in the $\mathcal{N} = 1$ superfield formulation has a non-polynomial form, as we do not work in the Wess–Zumino gauge. A gauge fixing term must anyway be added to the classical action. We shall use the Fermi-Feynman gauge, as it makes corrections to the propagators of the fundamental superfields vanish at order $g^2$ [33, 34, 35]. Actually a stronger result has been proved in these papers, namely the vanishing of the anomalous dimensions of the fundamental fields up to $O(g^4)$. With the Fermi-Feynman gauge choice the terms relevant for the calculation of the Green functions we are interested in are

$$S = \int d^4x \, d^2\theta d^2\bar{\theta} \left\{ V^a \Box V_a - \Phi_I^{a\dagger} \Phi_a^I - 2ig f_{abc} \Phi_I^{\dagger a} V^b \Phi^{Ic} + 2g^2 f_{abe} f_{ecd} \Phi_I^{\dagger a} V^b V^c \Phi^{Id} \right.$$
$$\left. - \frac{ig\sqrt{2}}{3!} f^{abc} \left[ \varepsilon_{IJK} \Phi_a^I \Phi_b^J \Phi_c^K \delta^{(2)}(\bar{\theta}) - \varepsilon^{IJK} \Phi_{aI}^\dagger \Phi_{bJ}^\dagger \Phi_{cK}^\dagger \delta^{(2)}(\theta) \right] + \ldots \right\}, \quad (1)$$

where $f_{abc}$ are the structure constants of the gauge group. As neither the cubic and quartic vector interactions nor the ghost terms will contribute to the calculations we will present in this paper, we have omitted them in eq. (1).

Since all superfields are massless, their propagators have an equally simple form in momentum and in coordinate space and thus we choose to work in the latter which is more suitable for the study of conformal field theories. In Euclidean coordinate space one finds

$$\langle \Phi_{Ia}^\dagger(x_i, \theta_i, \bar{\theta}_i) \Phi_b^J(x_j, \theta_j, \bar{\theta}_j) \rangle = \frac{\delta_I^{\,J} \delta_{ab}}{4\pi^2} e^{(\xi_{ii} + \xi_{jj} - 2\xi_{ji}) \cdot \partial_j} \frac{1}{x_{ij}^2}, \quad (2)$$

$$\langle V_a(x_i, \theta_i, \bar{\theta}_i) V_b(x_j, \theta_j, \bar{\theta}_j) \rangle = -\frac{\delta_{ab}}{8\pi^2} \frac{\delta^{(2)}(\theta_{ij}) \delta^{(2)}(\bar{\theta}_{ij})}{x_{ij}^2}, \quad (3)$$

where $x_{ij} = x_i - x_j$, $\theta_{ij} = \theta_i - \theta_j$, $\xi_{ij}^\mu = \theta_i^\alpha \sigma_{\alpha\dot\alpha}^\mu \bar\theta_j^{\dot\alpha}$.

The simplest protected (dimension two) CPO's

$$\mathcal{Q}_{\mathbf{20'}}^{(ij)} = \text{tr} \left( \varphi^i \varphi^j - \frac{\delta^{ij}}{6} \sum_k \varphi^k \varphi^k \right), \quad (4)$$

belong to the representation $\mathbf{20'}$ of $SU(4)$ and are the lowest component of the $\mathcal{N} = 4$ supercurrent multiplet.

In terms of $SU(3) \otimes U(1)$ the $\mathcal{Q}_{\mathbf{20'}}^{(ij)}$'s decompose in

$$\mathcal{C}^{IJ}(x) = \text{tr}(\phi^I(x)\phi^J(x)), \qquad \mathcal{C}_{IJ}^\dagger(x) = \text{tr}(\phi_I^\dagger(x)\phi_J^\dagger(x)) \quad (5)$$

and

$$\mathcal{V}_J^I = \text{tr}\left(e^{-2gc(x)} \phi_J^\dagger(x) e^{2gc(x)} \phi^I(x)\right) - \frac{\delta_J^I}{3} \text{tr}\left(e^{-2gc(x)} \phi_L^\dagger(x) e^{2gc(x)} \phi^L(x)\right), \quad (6)$$

where the exponentials ($c(x)$ is the lowest component of the vector superfield) are included to ensure gauge invariance and regularization of the operators is understood, *e.g.* by point-splitting (see below). Note that no normal-ordering is needed because the *vev*'s of all the above operators vanish, none of them being an $SU(4)$ singlet.

The $\mathcal{N}=4$ Konishi multiplet is a long multiplet of $SU(2,2|4)$ [36]. Its lowest component, $\mathcal{K}_\mathbf{1}$, is a scalar operator of (naive) conformal dimension $\Delta_0 = 2$, which is a singlet of the $SU(4)$ R-symmetry group. The (naive) definition of $\mathcal{K}_\mathbf{1}$ is

$$\mathcal{K}_\mathbf{1}(x)\Big|_\text{naive} = \frac{1}{2}\sum_{i=1}^{6} :\text{tr}(\varphi^i(x)\varphi^i(x)):, \qquad (7)$$

where the trace is over colour indices and the symbol :: stands for normal ordering. As usual, normal ordering means subtracting the operator *vev* or, in other words, requiring $\langle \mathcal{K}_\mathbf{1}\rangle = 0$. In terms of $\mathcal{N}=1$ superfields $\mathcal{K}_\mathbf{1}$ can be written in the form

$$\mathcal{K}_\mathbf{1}(x)\Big|_\text{formal} = \sum_{I=1}^{3} :\text{tr}(e^{-2gc(x)}\phi^\dagger_I(x)e^{2gc(x)}\phi^I(x)): . \qquad (8)$$

Since the operator $\mathcal{K}_\mathbf{1}$ has an anomalous dimension, $\gamma^\mathcal{K}(g^2)$, it will suffer a non-trivial renormalization. We assume (as is always the case in perturbation theory) that $\gamma^\mathcal{K}(g^2)$ is small and represented by the series expansion

$$\gamma^\mathcal{K}(g^2) = g^2\gamma^\mathcal{K}_1 + g^4\gamma^\mathcal{K}_2 + \ldots. \qquad (9)$$

From the results of refs. [37] and [20, 21], one gets for the first two coefficients of the expansion

$$\gamma^\mathcal{K}_1 = \frac{3N}{4\pi^2} \qquad (10)$$

$$\gamma^\mathcal{K}_2 = -\frac{3N^2}{16\pi^4}. \qquad (11)$$

It is convenient to regularize operators by point splitting. In particular for $\mathcal{K}_\mathbf{1}$ we write

$$\mathcal{K}_\mathbf{1}(x)\Big|_\text{reg} = a^\mathcal{K}(g^2)\sum_{I=1}^{3} :\text{tr}(e^{-2gc(x)}\phi^\dagger_I(x+\frac{\epsilon}{2})e^{2gc(x)}\phi^I(x-\frac{\epsilon}{2})):, \qquad (12)$$

where $\epsilon$ is an infinitesimal, but otherwise arbitrary, four-vector. Note that, due to our choice of gauge-fixing, there is no need to "point-split" the vector field in the exponents, because the $c$-field has vanishing propagator. Finally, the renormalized operator has the form

$$\mathcal{K}_\mathbf{1}(x)\Big|_\text{ren} = \lim_{\epsilon\to 0}\frac{a^\mathcal{K}(g^2)}{(\epsilon^2)^{\frac{1}{2}\gamma^\mathcal{K}(g^2)}}\sum_{I=1}^{3} :\text{tr}\left(e^{-2gc(x)}\phi^\dagger_I(x+\frac{\epsilon}{2})e^{2gc(x)}\phi^I(x-\frac{\epsilon}{2})\right):, \qquad (13)$$

where $a^\mathcal{K}(g^2)$ is a normalization factor that we choose of the form $a^\mathcal{K}(g^2) = 1 + g^2 a^\mathcal{K}_1 + g^4 a^\mathcal{K}_2 + \ldots$. Unlike the operators corresponding to symmetry generators (like the R-symmetry currents or the stress-energy tensor), the Konishi scalar $\mathcal{K}_\mathbf{1}$ has no intrinsic normalization, so we shall use this freedom in the normalization factor $a^\mathcal{K}(g^2)$ of eq. (13), to make the two-point function of $\mathcal{K}_\mathbf{1}$ depend on $g^2$ only through $\gamma^\mathcal{K}$. This is achieved at the order we work by setting $a_1^\mathcal{K} = 3N/8\pi^2$ (other coefficients would require higher order computations to be fixed). With this choice one gets

$$\langle \mathcal{K}_\mathbf{1}(x_1)\mathcal{K}_\mathbf{1}(x_2) \rangle = \frac{3(N^2-1)}{4(4\pi^2)^2} \frac{1}{(x_{12}^2)^{2+\gamma^\mathcal{K}(g^2)}}\,. \tag{14}$$

## 3 Comments on the UIR's of $SU(2,2|4)$

The unitary irreducible representations (UIR's) of $SU(2,2|4)$ have been classified in [38] [2]. A general UIR is specified by a set of quantum numbers comprising the dilation weight $\Delta$, the Lorentz spins $(j_1,j_2)$ and the Dynkin labels $[k,l,m]$ of the $SU(4)$ $R$-symmetry. There are three "unitary series", which are distinguished by different relations between the dilation weight and the other quantum numbers.

It has been known for some time that generic UIR's can be obtained by tensoring the so-called "singleton" representations [39] [3]. In [41] this has been elaborated in full detail using the technique of harmonic superspace [42]. Within this approach, in addition to the usual $\mathcal{N}=4$ super Minkowski space one introduces $4\times 4$ matrices, $u_r^A$, parameterizing the coset $SU(4)/U(1)^3$. We omit the details of the construction, but rather remark that these matrices should be contracted on all free $SU(4)$ indices in the constraints defining the "Grassmann analytic" superfields [43]

$$W^{[1\ldots k]}, \qquad 1 \leq k \leq 3 \tag{15}$$

which read

$$\begin{aligned} D_\alpha^A W^{[1\ldots k]} &= 0\,, & 1 \leq A \leq k\,, \\ \bar{D}_A^{\dot\alpha} W^{[1\ldots k]} &= 0\,, & k+1 \leq A \leq 4\,. \end{aligned} \tag{16}$$

In eq. (16) $D_\alpha^A = u_r^A D_\alpha^r$ are projected $\mathcal{N}=4$ supercovariant derivatives. The constraints express the fact that the $W$ superfields depend on half of the spinor coordinates, *i.e.* they are 1/2 BPS objects.

Besides the $W$'s, the list of singletons additionally includes $\mathcal{N}=4$ chiral superfields, which - unlike the former - may have either left or right handed spinor indices, but cannot carry a non-trivial $SU(4)$ representation. For the present purpose we only need to introduce the scalar chiral superfield, $\Psi$, which satisfies

$$\begin{aligned} \bar{D}_A^{\dot\alpha} \Psi &= 0\,, \\ D^{\alpha(A} D_\alpha^{B)} \Psi &= 0\,, \end{aligned} \tag{17}$$

where the second ('linear') constraint is a sort of 'field equation'. In the tensoring procedure it is assumed that the chiral superfield is 'on shell'.

---

[2]In (perturbative) $\mathcal{N}=4$ SYM theory only UIR's of $PSU(2,2|4)$ are actually relevant. They are characterized by the vanishing of the $U(1)_C$ central charge that extends $PSU(2,2|4)$ to $SU(2,2|4)$.

[3]Alternatively the UIR's may be built by using the oscillator method (see [40] and references therein) or they can be realized as "analytic tensor fields" on analytic superspace [23].

The statement that any UIR can be obtained as a product of singletons is formal in that none of the multiplets $W^1, W^{[123]}$ and $\Psi$ can be expressed in terms of elementary fields. On the contrary, $W^{[12]}$ is the fundamental $\mathcal{N} = 4$ SYM multiplet and its square is the supercurrent multiplet.

An important observation is that the component field content of the Konishi multiplet in the non-interacting theory is correctly reproduced by the product of a $\Psi$ field satisfying (17) with its complex conjugate:

$$\mathcal{K}_\mathbf{1}|_{g^0} = \Psi\bar\Psi \tag{18}$$

Using (17), one may verify that [4]

$$D^{\alpha(A} D_\alpha^{B)} \mathcal{K}_\mathbf{1}|_{g^0} = 0, \qquad \bar D_{\dot\alpha(A} \bar D_{B)}^{\dot\alpha} \mathcal{K}_\mathbf{1}|_{g^0} = 0. \tag{19}$$

By acting with the products of $D^{\alpha(A} D_\alpha^{B)}$ and $\bar D_{\dot\alpha(A} \bar D_{B)}^{\dot\alpha}$ on $\mathcal{K}_\mathbf{1}|_{g^0}$ and performing some $D$-algebra, one realizes that both the singlet and the **15** components of the current

$$K_{\mu\,B}^A = \bar\sigma_\mu^{\dot\alpha\alpha} [D_\alpha^A, \bar D_{\dot\alpha\,B}] \mathcal{K}_\mathbf{1}|_{g^0,\theta,\bar\theta=0} \tag{20}$$

are conserved. There are similar constraints on some of the higher components. Also, each of the linear constraints (19) separately implies the absence of some component fields in the supermultiplet.

In the interacting theory the Konishi multiplet has an anomalous dimension, so we may formally write

$$\mathcal{K}_\mathbf{1} = (\Psi\bar\Psi)^{(1+\gamma)}. \tag{21}$$

For $\gamma \neq 0$ it does not satisfy any differential constraint, consistently with it being a long multiplet.

Next, we focus on operators of naive dimension 4 in the **20'**. Operators in this representation can be either single or double trace composites of the fields in the fundamental $\mathcal{N} = 4$ SYM multiplet. Independently of their actual expression, the formalism of [41] allows to determine the component field content multiplets of this kind by representing it in the form

$$\mathcal{O}_{\mathbf{20'}} = (\Psi\bar\Psi)^{(1+\gamma)} (W^{[12]})^2. \tag{22}$$

Eq. (22) shows that it has the same field content as a product of the Konishi and supercurrent multiplets.

If and only if $\gamma = 0$, the operator satisfies differential constraints which are in the intersection of the conditions (16) and (19), *i.e.*

$$\begin{aligned} D^{\alpha A} D_\alpha^B \mathcal{O}_{\mathbf{20'}} &= 0, & 1 \leq A, B \leq 2, \\ \bar D_{\dot\alpha A} \bar D_B^{\dot\alpha} \mathcal{O}_{\mathbf{20'}} &= 0, & 3 \leq A, B \leq 4. \end{aligned} \tag{23}$$

Note that the derivatives $D^1, D^2, \bar D_3, \bar D_4$ mutually anticommute, so that we can never derive a condition containing a space time derivative from them. Hence, even in the non-interacting case, such multiplets do not contain conserved tensor currents. The constraints (23) merely express the absence of some component fields.

---

[4]In the interacting theory the situation is more complicated, as the r.h.s. of the equations in (19) do not vanish [25]. For instance, the first becomes $D^{\alpha(A} D_\alpha^{B)} \mathcal{K}_\mathbf{1} \propto g\mathrm{tr}([W^{AC}, W^{BD}] \bar W_{CD})$. These considerations are at the basis of the derivation of the Konishi anomaly [44].

If the conditions (23) happen to be enforced by some mechanism, *e.g.* because the field occurs in the OPE of two supercurrent multiplets [22], then the operator will have protected dimension. This is exactly what happens for the double-trace operator $\mathcal{D}_{\mathbf{20'}}$. Its explicit expression, obtained in [18], is given for convenience in eq. (31). It may be argued that the reason why this operator is protected is to be ascribed to the fact that it obeys constraints not involving SYM covariant superderivatives [25].

## 4 Scalar operators in the $\mathbf{20'}$ representation

For a sufficiently large number of colours, $N \geq 4$, there are 6 distinct scalar primary operators of naive scale dimension $\Delta_0 = 4$ in the real representation $\mathbf{20'}$ of $SU(4)$. Two of them $\mathcal{K}^{\pm}_{\mathbf{20'}}$ belong to the Konishi multiplet. At leading order in $g$ they correspond to generalized Yukawa couplings

$$\mathcal{K}^{ij}_{\mathbf{20'}} = t^{(i}_{AB} Tr(\varphi^{j)} [\lambda^A, \lambda^B]) + \ldots, \qquad (24)$$

which are bilinear in the fermions. Hence they do not contribute at tree level and, in general, to the leading logarithms, $g^{2n}(\log x_{12}^2)^n$, in the correlation functions that we shall analyze. The other four operators are quartic in the fundamental scalars (at leading order in $g$) and naively can be written in the form [5]

$$\mathcal{O}^{11}_\ell(x)|_{\text{naive}} = \sum_{a,b,c,d} : \phi^1_a(x)\phi^1_b(x) \sum_{I=1}^{3} \left( \phi^I_c(x)\phi^\dagger_{I,d}(x) + \phi^I_d(x)\phi^\dagger_{I,c}(x) \right) : X^{abcd}_\ell \quad , \qquad (25)$$

where the index $\ell = 1, 2, 3, 4$ labels the four different operators characterized by the four colour tensors $X^{abcd}_\ell$, given below (eqs. (30) to (35)).

To properly define renormalized operators one has to make three modifications to the naive formula (25). First, in order to ensure gauge invariance the fundamental scalars have to be replaced by

$$\begin{aligned} \phi^\dagger_I(x) &\to \mathrm{e}^{-gc(x)} \phi^\dagger_I(x) \mathrm{e}^{gc(x)}, \\ \phi^I(x) &\to \mathrm{e}^{gc(x)} \phi^I(x) \mathrm{e}^{-gc(x)}, \end{aligned} \qquad (26)$$

respectively. Second, one has to regularize the operator. We choose to do this by point splitting, separating the arguments of the four scalars in (25) by a small distance. There are several, essentially equivalent, ways to do this, the most compact one is to position the four scalars at the vertices of a tetrahedron so that all the separations have equal length which we shall denote by $\epsilon$. As remarked after eq.(12), there is no need to point split the different gauge exponents in (26). Still one has to separate their arguments from the arguments of the fundamental scalars. We decided to put them all at the center of the tetrahedron. Finally, as in the case of $\mathcal{K}$, operators that acquire anomalous dimension have to be renormalized. Within the regularization prescription we chose this amounts to writing

$$\mathcal{O}_\ell^{11}(x)|_{\text{ren}} = \lim_{\epsilon \to 0} \frac{a^\ell(g^2)}{(\epsilon^2)^{\frac{1}{2}\gamma^{O_\ell}(g^2)}} \left(\mathcal{O}_\ell^{11}(x)|_{\text{reg}} + \ldots\right), \qquad (27)$$

where dots in the r.h.s. denote possible subleading (in $g$) mixings which will not contribute at the order we work. Here $\gamma^{O_\ell}(g^2)$ is the anomalous dimension of the operator $O_\ell$ for which as before we assume the power series expansion

$$\gamma^{O_\ell}(g^2) = g^2 \gamma_1^{O_\ell} + g^4 \gamma_2^{O_\ell} + \ldots, \qquad (28)$$

while $a^\ell(g^2)$ is a finite renormalization of the form

$$a^\ell(g^2) = 1 + g^2 a_1^\ell + g^4 a_2^\ell + \ldots, \qquad (29)$$

which depends on the regularization prescription.

---
[5]We shall omit the $\mathbf{20'}$ label from now on and choose to deal with one of the 20 components of the multiplet that we take to be the one belonging to the representation $\mathbf{6}_{+2}$ of $SU(3) \otimes U(1)$ in the decomposition $\mathbf{20'} \to \mathbf{8}_0 \oplus \mathbf{6}_{+2} \oplus \mathbf{6}^*_{-2}$, namely $\mathcal{O}^{11}$.

Let us now list the four (for $N \geq 4$) colour tensors $X_\ell^{abcd}$. The protected operator, $\mathcal{D}$, is double trace and corresponds to the tensor

$$X_\mathcal{D}^{abcd} = \frac{1}{2}\delta^{ac}\delta^{bd} - \frac{1}{6}\delta^{ab}\delta^{cd}. \qquad (30)$$

Note that $\mathcal{D}$ can be expressed in terms of only the protected $\Delta = 2$ CPO's $\mathcal{Q}^{(ij)}$ of eq. (4) as

$$\mathcal{D}^{(ij)} = \sum_k \mathcal{Q}^{(ik)}\mathcal{Q}^{(jk)} - \frac{\delta^{ij}}{6} \sum_{k,\ell} \mathcal{Q}^{(k\ell)}\mathcal{Q}^{(k\ell)}. \qquad (31)$$

We have to choose an appropriate basis for the remaining three operators, which we denote by $\mathcal{M}$, $\mathcal{P}$ and $\mathcal{L}$. This choice is at this point purely conventional, since, as we shall demonstrate, the correct operators, *i.e.* those that have well defined (anomalous) dimensions, will turn out to be complicated linear combinations of the former. We denote by $\mathcal{M}$ the double trace operator corresponding to the colour tensor

$$X_\mathcal{M}^{abcd} = \frac{1}{2}\delta^{ab}\delta^{cd}, \qquad (32)$$

Note that $\mathcal{M}$ is the product of the lowest chiral primary operator $\mathcal{Q}^{(ij)}$ of eq. (4) and the lowest operator in the Konishi supermultiplet, $\mathcal{K}_1$, given in eq. (7), *i.e.*

$$\mathcal{M}^{(ij)} = \mathcal{Q}^{(ij)} \mathcal{K}_1. \qquad (33)$$

We denote by $\mathcal{P}$ the single trace operator corresponding to

$$X_\mathcal{P}^{abcd} = -\frac{1}{2} \sum_e f_{ace} f_{bde}. \qquad (34)$$

Note that for $SU(2)$ $\mathcal{P}$ is a linear combination of $\mathcal{D}$ and $\mathcal{M}$. The fourth operator, $\mathcal{L}$, is constructed by saturating colour indices with the (traceless) quartic Casimir operator of $SU(N)$

$$\begin{aligned}X_\mathcal{L}^{abcd} &= \mathrm{Tr}(T^a T^b T^c T^d) + \text{permutations of } b, c, d \\ &- \frac{1}{2}\frac{2N^2 - 3}{N(N^2 + 1)}\left(\delta^{ab}\delta^{cd} + \delta^{ac}\delta^{bd} + \delta^{ad}\delta^{bc}\right),\end{aligned} \quad (35)$$

where $T^a$ are the generators of $SU(N)$ in the fundamental representation. $X_\mathcal{L}$ vanishes both for $SU(2)$ and $SU(3)$.

From this general discussion it is clear that one has to treat separately the two low rank cases of $SU(2)$ and $SU(3)$, since the number of independent operator varies with $N$.

## 5 Orthogonalization of the scalar operators

We shall consider only tree-level and order $g^2$ constraints, since higher loop quantities will depend also on terms omitted in equation (27). The correctly renormalized operators (*i.e.* those having well defined anomalous dimensions) must satisfy the following three orthogonality properties.

1. They have to be orthogonal to (*i.e.* they must have vanishing two-point functions with) the two operators belonging to the Konishi multiplet, $\mathcal{K}_{\mathbf{20'}}^\pm$. This is automatic both at tree level and for the leading logarithms at order $g^2$, because $\mathcal{K}_{\mathbf{20'}}^\pm$ are bilinear in the fermions at leading order.

2. They have to be orthogonal to the protected operator $\mathcal{D}$. An explicit calculation shows that both at tree level and at order $g^2$ the operator $\mathcal{L}$ already enjoys this property, while for the other two one needs to introduce the definitions

$$\widehat{\mathcal{M}} = \mathcal{M} - \frac{6}{3N^2 - 2}\mathcal{D} \quad (36)$$

$$\widehat{\mathcal{P}} = \mathcal{P} + \frac{2}{3}\frac{N}{N^2 - 2}\mathcal{M} - \frac{N}{N^2 - 2}\mathcal{D}. \quad (37)$$

Indeed any linear combination of $\widehat{\mathcal{M}}$ and $\widehat{\mathcal{P}}$ is orthogonal (in the sense explained above) to $\mathcal{D}$. Our choice is such that they are mutually orthogonal at tree level for any $N$ and the operator $\widehat{\mathcal{P}}$ vanishes for $SU(2)$. Thus for $SU(2)$ the only relevant operator is $\widehat{\mathcal{M}}$. From the presence of logarithmic terms at short distances in the expansion of the four-point function $\langle \mathcal{C}^{11}(x_1)\mathcal{C}_{11}^\dagger(x_2)\mathcal{K}_1(x_3)\mathcal{K}_1(x_4)\rangle$ at order $g^2$ it follows that $\widehat{\mathcal{M}}$ has a non-vanishing one-loop anomalous dimension. The precise value was computed in [18] with the result

$$\gamma_1^{\widehat{\mathcal{M}}} = 5 \times \frac{Ng^2}{4\pi^2}, \quad N = 2. \quad (38)$$

3. Finally they have to be mutually orthogonal both at tree level and at one loop. In other words we have to define new operators

$$\begin{aligned}\mathcal{O}_1 &= \alpha_{1\mathcal{M}}\widehat{\mathcal{M}} + \alpha_{1\mathcal{P}}\widehat{\mathcal{P}} + \alpha_{1\mathcal{L}}\mathcal{L} \\ \mathcal{O}_2 &= \alpha_{2\mathcal{M}}\widehat{\mathcal{M}} + \alpha_{2\mathcal{P}}\widehat{\mathcal{P}} + \alpha_{2\mathcal{L}}\mathcal{L} \\ \mathcal{O}_3 &= \alpha_{3\mathcal{M}}\widehat{\mathcal{M}} + \alpha_{3\mathcal{P}}\widehat{\mathcal{P}} + \alpha_{3\mathcal{L}}\mathcal{L}\end{aligned} \quad (39)$$

and require

$$\langle \mathcal{O}_i(x)\mathcal{O}_j(y)\rangle = 0 \quad \text{if } i \neq j \tag{40}$$

both at tree level and at order $g^2$.

It follows from (27) that the order $g^2$ correction to the anomalous dimension of the operator $\mathcal{O}_i$ is equal to the ratio of the coefficient of the $\ln(x-y)^2$ term at order $g^2$ and the tree level normalization of the two-point function of $\mathcal{O}_i$, i.e.

$$\gamma_1^{\mathcal{O}_i} = -\frac{\langle \mathcal{O}_i(x)\mathcal{O}_i(y)\rangle|_{g^2,\ln(x-y)^2}}{\langle \mathcal{O}_i(x)\mathcal{O}_i(y)\rangle|_0} \,. \tag{41}$$

Hence it suffices to solve the orthogonality relations (40) to get the explicit expressions of the $\mathcal{O}_i$'s anomalous dimensions.

Before proceeding to this rather long calculation let us make two comments. First, we notice that the order $g^2$ contributions to these two-point functions come from two different types of diagrams, corresponding to either chiral or vector internal lines. Both types of diagrams can be expressed in terms of the massless box integral (see the Appendix). At any order in $g^2$ each orthogonality relation (40) leads to one equation for the coefficients $\alpha$ in (39). These equations ensure the cancellation of all $\ln(\epsilon)$-singularities. At this point, by a suitable choice of the coefficients $a_1^\ell$ in (29) one can also cancel the finite corrections at the same order. Second, since the overall normalization of the operators $\mathcal{O}_i$ is arbitrary, 3 out of the 9 parameters in (39) can be chosen without loss of generality to be equal to 1. Thus the orthogonality constraint leads to 6 quadratic equations for 6 variables. A careful analysis shows that up to permutation symmetry there is only one acceptable (i.e. real) solution.

The two-point function of the protected operator $\mathcal{D}$ is given by its tree-level value

$$\langle \mathcal{D}\mathcal{D}\rangle|_0 = \langle \mathcal{D}(x_1)\mathcal{D}(x_2)\rangle = \frac{10}{9}(N^2-1)(3N^2-2)\mathcal{I}_0(x_{12})\,, \tag{42}$$

where

$$\mathcal{I}_0(x_{12}) = \frac{1}{(4\pi^2)^4(x_{12}^2)^4}\,. \tag{43}$$

For the other operators at tree level we find

$$\langle \widehat{\mathcal{M}}\widehat{\mathcal{M}}\rangle|_0 = \frac{18(N^2-1)(N^2-2)(N^2+1)}{(3N^2-2)} \mathcal{I}_0(x_{12})\,, \tag{44}$$

$$\langle \widehat{\mathcal{P}}\widehat{\mathcal{P}}\rangle|_0 = \frac{3N^2(N^2-1)(N^2-4)}{(N^2-2)} \mathcal{I}_0(x_{12})\,, \tag{45}$$

$$\langle \mathcal{L}\mathcal{L}\rangle|_0 = \frac{15(N^2-1)(N^2-4)(N^2-9)}{(N^2+1)} \mathcal{I}_0(x_{12})\,, \tag{46}$$

while the off-diagonal functions vanish by construction.

At order $g^2$ we obtain

$$\langle \widehat{\mathcal{M}}\widehat{\mathcal{M}}\rangle|_{g^2} = 18N(N^2-1)(N^2+1)\,\mathcal{I}(x_{12})\,, \tag{47}$$

$$\langle \widehat{\mathcal{P}}\widehat{\mathcal{M}}\rangle|_{g^2} = -\frac{18N^2(N^2-1)(N^2-4)}{(N^2-2)} \mathcal{I}(x_{12})\,, \tag{48}$$

$$\langle \mathcal{L}\widehat{\mathcal{M}}\rangle|_{g^2} = 0, \tag{49}$$

$$\langle \widehat{\mathcal{P}}\widehat{\mathcal{P}}\rangle|_{g^2} = \frac{2N^3(N^2-1)(N^2-4)(5N^2-16)}{(N^2-2)^2}\mathcal{I}(x_{12}), \tag{50}$$

$$\langle \mathcal{L}\widehat{\mathcal{P}}\rangle|_{g^2} = -\frac{5N(N^2-1)(N^2-4)(N^2-9)}{(N^2+1)}\mathcal{I}(x_{12}), \tag{51}$$

$$\langle \mathcal{L}\mathcal{L}\rangle|_{g^2} = \frac{25N(N^2-1)(N^2-4)(N^2-9)}{(N^2+1)}\mathcal{I}(x_{12}), \tag{52}$$

where $\mathcal{I}(x_{12})$ is proportional to the short distance limit of the box integral. Using eq. (91) in the Appendix, one finds

$$\mathcal{I}(x_{12}) = -\frac{1}{(4\pi)^5(x_{12}^2)^4}\left(\ln(x_{12}^2)+1\right). \tag{53}$$

Let us start by analyzing the generic case $N \geq 4$. The cases $N=3$ and $N=2$ are much simpler and we shall discuss them in Sect. 7. Since, as we said, all coefficients in (39) turn out to be non-vanishing, without loss of generality we shall set $\alpha_{i\mathcal{P}} = 1$. With this choice the orthogonality relations lead to the equation

$$\alpha_{i\mathcal{M}} = \frac{\zeta_i}{(N^2+1)(N^2-2)}, \tag{54}$$

where $\zeta_i$ are the three roots of the cubic equation

$$8N\zeta^3 + (-N^6 + 2N^4 + 68N^2 - 40)\zeta^2 + (-3N^9 - 16N^7 + 132N^5 - 80N^3)\zeta$$
$$-84N^{10} + 64N^4 + 244N^8 - 224N^6 + 9N^{12} = 0. \tag{55}$$

For the coefficients $\alpha_{i\mathcal{L}}$ we get

$$\alpha_{i\mathcal{L}} = -\frac{24}{5}\alpha_{i\mathcal{M}}^2\frac{(N^2+1)^3}{(3N^2-2)(N^2-4)(N^2-9)}$$
$$+ \frac{3}{5}\alpha_{i\mathcal{M}}\frac{(N^2-2N+2)(N^2+2N+2)(N^4-19N^2+10)(N^2+1)}{N(N^2-2)(3N^2-2)(N^2-4)(N^2-9)}$$
$$+ \frac{4}{5}\frac{(N^2+1)^2 N^2}{(N^2-2)^2(N^2-9)}. \tag{56}$$

Inserting all these formulae into equations (41), we obtain for the order $g^2$ corrections to the anomalous dimensions of the operators $\mathcal{O}_i$ the following expressions

$$\gamma_1^{\mathcal{O}_i} = \frac{g^2 N}{4\pi^2}\frac{8\zeta_i^2 N - (N^6 - 2N^4 - 68N^2 + 40)\zeta_i + 6N^9 - 64N^7 + 184N^5 - 96N^3}{N^3(N^2-4)(3N^2-2)(N^2-2)}, \tag{57}$$

where again the $\zeta_i$'s are the three roots of the cubic equation (55).

In Fig. 1 we plot the ratios $\gamma_1^{\mathcal{O}_i}/\gamma_1^{\mathcal{K}}$, $i=1,2,3$, as functions of $N$, where $\gamma_1^{\mathcal{K}}$ is the order $g^2$ coefficient of the Konishi multiplet anomalous dimension, displayed in eq. (10).

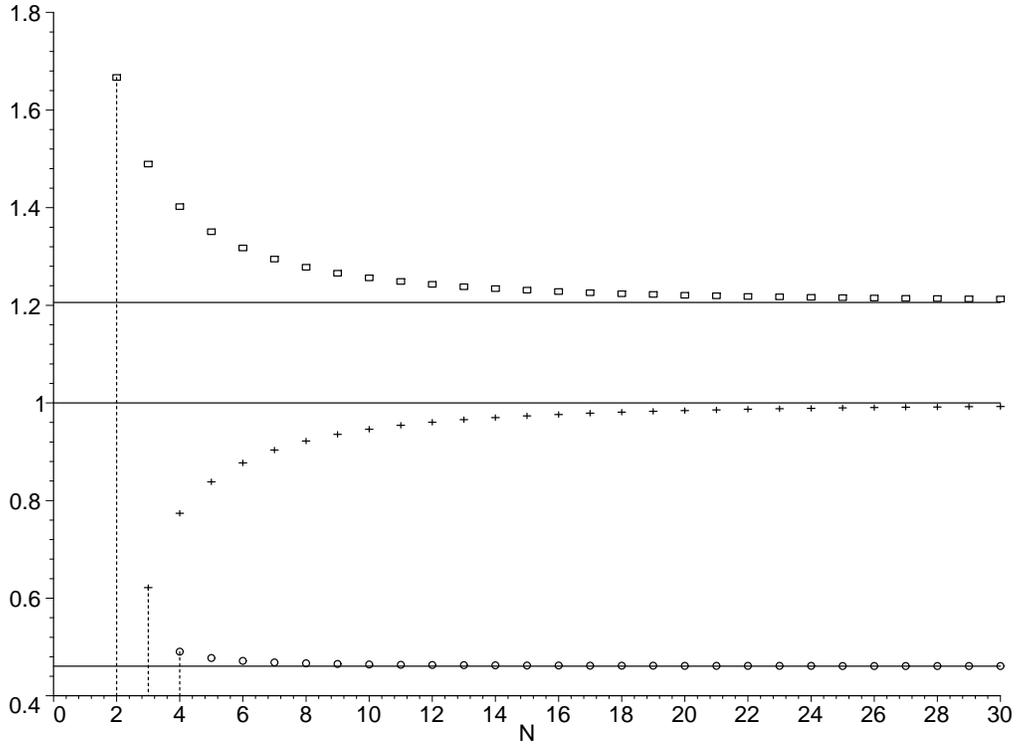

Figure 1: The ratios $\gamma_1^{\mathcal{O}_i}/\gamma_1^{\mathcal{K}}$, $i = 1, 2, 3$, as functions of $N$. $\gamma_1^{\mathcal{K}}$ is the order $g^2$ coefficient of the Konishi multiplet anomalous dimension, displayed in eq. (10).

## 5.1 The $N \to \infty$ limit

We would like to end this section by briefly discussing the peculiar properties of the above orthogonalization procedure in the $N \to \infty$ limit. To this purpose we notice that

1) multiple trace operators with different number of traces are mutually orthogonal in this limit.

2) in the space of multiple $n$-trace operators one can choose a basis of "product" operators of the form

$$\prod_{j=1...n} \tilde{\mathcal{O}}_j \, , \tag{58}$$

where the $\tilde{\mathcal{O}}_j$'s are mutually orthogonal single trace operators of well defined anomalous dimensions, $\gamma(\tilde{\mathcal{O}}_j)$.

Note that both double trace operators defined above, *i.e.* $\mathcal{D}$ of eq. (31) and $\mathcal{M}$ of eq. (33), have the form (58) (the $SU(4)$ trace subtraction in $\mathcal{D}$ does not affect the following argument). Since for large $N$ the two-point function of a multiple trace "product" operator is dominated by the most disconnected part (i.e. by the products of the two point functions of the constituent operators), different "product" operators will be orthogonal for large $N$ due to the orthogonality of the $\tilde{\mathcal{O}}_j$'s. Moreover, it follows that in the $N \to \infty$ limit the anomalous dimension of the "product" operator is the sum of the anomalous dimensions of its constituent operators, namely

$$\lim_{N \to \infty} \gamma(\prod_{j=1...n} \tilde{\mathcal{O}}_j) = \sum_{j=1...n} \lim_{N \to \infty} \gamma(\tilde{\mathcal{O}}_j) \, . \tag{59}$$

Let us now return to the case at hand. From Fig. 1 one can see that for large $N$ the anomalous dimension of one of the three operators tends to the anomalous dimension of the Konishi supermultiplet, $\gamma_1^{\mathcal{K}}$. An explicit calculation confirms that this operator is indeed $\mathcal{M}$ (see eq. (33)). The anomalous dimensions of the other two operators, which in line with the above discussion for large $N$ will be dominantly single trace ones, tend to

$$\lim_{N \to \infty} \gamma_\pm = \frac{1}{6}(5 \pm \sqrt{5})\gamma_1^{\mathcal{K}} \, . \tag{60}$$

The corresponding colour tensors $X^{abcd}$ entering eq. (25) are in the $N \to \infty$ limit

$$X_\pm^{abcd} = \sqrt{\sqrt{5} \pm 1} \, \text{Tr}(T^a T^b T^c T^d) \mp \sqrt{\sqrt{5} \mp 1} \, \text{Tr}(T^a T^c T^b T^d) \, . \tag{61}$$

## 6 The four point function $\langle \mathcal{C}^{11} \mathcal{C}_{11}^\dagger \mathcal{K}_1 \mathcal{K}_1 \rangle$ at order $g^4$

Given the complexity of the results we got in the previous section, it is important to rederive them from an alternative point of view, *i.e.* by performing an OPE analysis of appropriate four-point functions.

Since the only operator of dimension 4 in the $\mathbf{20'}$ that appears in the OPE of two CPO's $\mathcal{Q}$ is the protected operator $\mathcal{D}_{\mathbf{20'}}$ (see eqs. (25) and (30)) [26], little mileage can be gained by studying only the correlation function of four $\mathcal{Q}$'s. The next simplest four-point functions involve the lowest component of the Konishi multiplet. Correlation functions involving two Konishi operators like $\langle \mathcal{Q}\mathcal{Q}\mathcal{K}\mathcal{K}\rangle$ meet all the necessary requirements. This kind of correlators have been calculated at lowest order in $g^2$ in [18], where it was also shown that they receive no instanton corrections.

Since there is only one $SU(4)$ tensor structure, without loss of generality we can use the following representative for this correlator

$$G(x_1, x_2, x_3, x_4) = \langle \mathcal{C}^{11}(x_1)\mathcal{C}^\dagger_{11}(x_2)\mathcal{K}_1(x_3)\mathcal{K}_1(x_4)\rangle \,, \tag{62}$$

which with the help of the identity (no summation on $J$)

$$: tr(\phi^\dagger_J(x)\phi^J(x)) := \mathcal{V}^J_J(x) + \frac{1}{3}\mathcal{K}_1(x). \tag{63}$$

can be written also as

$$G(x_1, x_2, x_3, x_4) = \tag{64}$$
$$9\langle \mathcal{C}^{11}(x_1)\mathcal{C}^\dagger_{11}(x_2) : tr(\phi^\dagger_2(x_3)\phi^2(x_3)) :: tr(\phi^\dagger_3(x_4)\phi^3(x_4)) :\rangle -$$
$$3\langle \mathcal{C}^{11}(x_1)\mathcal{C}^\dagger_{11}(x_2)\mathcal{K}_1(x_3)\mathcal{V}^3_3(x_4)\rangle - 3\langle \mathcal{C}^{11}(x_1)\mathcal{C}^\dagger_{11}(x_2)\mathcal{V}^2_2(x_3)\mathcal{K}_1(x_4)\rangle -$$
$$9\langle \mathcal{C}^{11}(x_1)\mathcal{C}^\dagger_{11}(x_2)\mathcal{V}^2_2(x_3)\mathcal{V}^3_3(x_4)\rangle + \langle \mathcal{C}^{11}(x_1)\mathcal{C}^\dagger_{11}(x_2)\rangle\langle\mathcal{K}_1(x_3)\mathcal{K}_1(x_4)\rangle \,.$$

For the purposes of our analysis only the first (connected) term in the r.h.s. of eq. (64) has really to be computed at order $g^4$. In fact in the OPE limit $x_{13} \to 0$, the leading $\ln^2(x^2_{13})$ contributions in the $\mathbf{20'}$ channel that comes from the other correlators in eq. (64) can be obtained simply by using tree level and order $g^2$ data.

In order to simplify the calculation it is convenient to make use of conformal invariance to map one of the coordinate points to infinity, while at the same time appropriately rescaling the corresponding field operator. Introducing the abbreviated notation

$$\mathcal{C}^\dagger_{11}(\infty) = \lim_{x_2 \to \infty} x_2^4 \, \mathcal{C}^\dagger_{11}(x_2) \,, \tag{65}$$

we find after a lengthy calculation (which is sketched in the Appendix)

$$\langle \mathcal{C}^{11}(x_1)\mathcal{C}^\dagger_{11}(\infty) : tr(\phi^\dagger_2(x_3)\phi^2(x_3)) :: tr(\phi^\dagger_3(x_4)\phi^3(x_4)) :\rangle|_{g^4} =$$

$$\frac{N^2(N^2-1)}{16(4\pi^2)^6}\left[\frac{1}{2x^4_{34}}B\left(\frac{x^2_{13}}{x^2_{34}},\frac{x^2_{14}}{x^2_{34}}\right)\left(\frac{x^2_{13}}{x^2_{14}}+\frac{x^2_{14}}{x^2_{13}}-\frac{x^2_{34}}{x^2_{13}}-\frac{x^2_{34}}{x^2_{14}}-2\right) + \right.$$
$$\frac{1}{2x^2_{13}x^2_{34}}\ln\left(\frac{x^2_{13}}{x^2_{14}}\right)\ln\left(\frac{x^2_{14}}{x^2_{34}}\right) + \frac{1}{2x^2_{14}x^2_{34}}\ln\left(\frac{x^2_{14}}{x^2_{13}}\right)\ln\left(\frac{x^2_{13}}{x^2_{34}}\right) +$$
$$\frac{1}{x^2_{13}x^2_{34}}\left(\ln(x^2_{34})+1\right)\left(\ln\left(\frac{x^2_{13}x^2_{34}}{x^2_{14}}\right)+1\right) +$$
$$\frac{1}{x^2_{14}x^2_{34}}\left(\ln(x^2_{34})+1\right)\left(\ln\left(\frac{x^2_{14}x^2_{34}}{x^2_{13}}\right)+1\right) +$$
$$\left.\frac{4}{x^4_{34}}B\left(\frac{x^2_{13}}{x^2_{34}},\frac{x^2_{14}}{x^2_{34}}\right)\left(\ln(x^2_{34})+1\right) + \frac{1}{x^2_{13}x^2_{14}}\left(\ln(x^2_{13})+1\right)\left(\ln(x^2_{14})+1\right)\right] \,, \tag{66}$$

where $B$ is the box integral given in eq. (88).

From eq.(66) we can now determine the coefficient of the leading logarithmic terms in $G$ associated to the scalar operators of (naive) dimension $\Delta_0 = 4$. This requires to identify also all the other operators with (naive) dimension smaller or equal to 4 and to subtract their contributions. In particular for the coefficient of $\ln^2(x_{13}^2)$ in (66), we find the factor $-N^2(N^2-1)/(32(4\pi^2)^6)$. Looking at the other correlators in eq. (64), we see that in the functions $\langle \mathcal{C}^{11}\mathcal{C}_{11}^\dagger \mathcal{K}_1 \mathcal{V}_3^3 \rangle$ and $\langle \mathcal{C}^{11}\mathcal{C}_{11}^\dagger \mathcal{V}_2^2 \mathcal{K}_1 \rangle$ only the protected operators of dimension $\Delta_0 = 2$ and $\Delta_0 = 4$ can appear in the relevant intermediate channel, while in $\langle \mathcal{C}^{11}\mathcal{C}_{11}^\dagger \mathcal{V}_2^2 \mathcal{V}_3^3 \rangle$ the component $\mathcal{K}_{84}$ of the Konishi multiplet alone contributes. Finally the disconnected parts in eq. (64) give no contribution to the $\ln^2(x_{13}^2)$ term. The explicit calculation of the residual coefficients after subtracting the known contribution of all these operators confirms that for $N = 2$ at level $\Delta_0 = 4$ in the representation $\mathbf{20'}$ of $SU(4)$ there is only one unprotected scalar operator, namely $\widehat{\mathcal{M}}$, having anomalous dimension given by eq. (38). This fact will allow us to compute in the next section the anomalous dimensions of all the scalar operators appearing at the same level in the representation $\mathbf{20'}$ of $SU(4)$ for $N > 2$.

# 7 A second derivation of the order $g^2$ anomalous dimension of the $\mathbf{20'}$ operators

In this section we present an alternative computation of the anomalous dimensions of the scalar operators with naive dimension $\Delta_0 = 4$ belonging to the representation $\mathbf{20'}$ of $SU(4)$, that does not require their explicit construction in terms of fundamental fields.

The basic assumption is that, according to conformal invariance, logarithmic terms exponentiate to a power law. This has been confirmed in all order $g^2$- and $g^4$- calculations performed so far.

Another important ingredient is the observation that four-point functions of single trace operators quadratic in the fundamental fields (like the protected operators $\mathcal{C}^{11}$, $\mathcal{C}_{11}^\dagger$ as well as the lowest component of the Konishi multiplet $\mathcal{K}$) have a particularly simple polynomial behaviour in $N$. Indeed, expanding the interaction Lagrangian in powers of $g$ one observes that the coefficient of $g^{2n}$ contains the product of exactly $2n$ $SU(N)$ structure constants $f^{abc}$, hence the connected part of the perturbative amplitude can be represented as a single trace of $2n$ $SU(N)$ matrices in the adjoint representation. It is a rather long, but straightforward, computation to evaluate all the traces we shall need (i.e. those that can appear in perturbation theory up to order $g^{10}$). Below we list the types of $N$ behaviour that one encounters. The coefficients with which these power behaviours appear depend on the particular trace one is considering and will be of no interest here. One gets

- order $g^2$: only $N(N^2-1)$;
- order $g^4$: only $N^2(N^2-1)$;
- order $g^6$: only $N^3(N^2-1)$;

- order $g^8$: linear combination of $N^4(N^2-1)$ and $N^2(N^2-1)$;
- order $g^{10}$: linear combination of $N^5(N^2-1)$ and $N^3(N^2-1)$

In the computation of the four-point function $G$ there are both connected and disconnected contributions (see eq. (64)). The disconnected piece is effectively double trace, so it is multiplied by one more factor of $N^2-1$ with respect to the connected term.

We consider the double OPE $x_{13}\to 0$, $x_{24}\to 0$, i.e. the exchange of operators between the products $\mathcal{C}^{11}(x_1)\mathcal{K}_1(x_3)$ and $\mathcal{C}^\dagger_{11}(x_2)\mathcal{K}_1(x_4)$, and concentrate on the leading $\ln(x_{34}^2)$ behaviour only. To be more precise we shall analyze the coefficient of the terms $g^{2n}\ln^n(x_{34}^2)$ in the correlator (62). A convenient normalization is to extract a common factor equal to

$$\frac{(-\gamma_1^\mathcal{K})^n}{n!}\,\frac{\langle\mathcal{C}^{11}\mathcal{C}^\dagger_{11}\rangle_0\,\langle\mathcal{K}_1\mathcal{K}_1\rangle_0}{(N^2-1)}\,,\tag{67}$$

where

$$\langle\mathcal{C}^{11}\mathcal{C}^\dagger_{11}\rangle_0 = \frac{(N^2-1)}{2(4\pi^2)^2}\,,\tag{68}$$

$$\langle\mathcal{K}_1\mathcal{K}_1\rangle_0 = \frac{3(N^2-1)}{4(4\pi^2)^2}\tag{69}$$

are the tree level normalizations of the two-point functions of the operators $\mathcal{C}^{11}$ and $\mathcal{K}_1$, respectively, while $\gamma_1^\mathcal{K} = 3N/(4\pi^2)$ is the order $g^2$ correction to the anomalous dimension of the Konishi multiplet (eq. (10)).

With these choices the contribution from the disconnected diagrams to the $g^{2n}\ln^n(x_{34}^2)$ term is $N^2-1$ for any value of $n$. It is convenient to "measure" anomalous dimensions in units of the anomalous dimension of the Konishi multiplet, by defining

$$\eta_i = \frac{\gamma_1^{\mathcal{O}_i}}{\gamma_1^\mathcal{K}}\tag{70}$$

and to introduce the ratio

$$F_i = \frac{N^2-1}{\langle\mathcal{C}^{11}\mathcal{C}^\dagger_{11}\rangle_0\,\langle\mathcal{K}_1\mathcal{K}_1\rangle_0}\frac{\langle\mathcal{C}^{11}\mathcal{K}_1\mathcal{O}_i\rangle_0\,\langle\mathcal{C}^\dagger_{11}\mathcal{K}_1\mathcal{O}_i^\dagger\rangle_0}{\langle\mathcal{O}_i\mathcal{O}_i^\dagger\rangle_0}\tag{71}$$

to normalize the contributions of the different operators to the four-point function (64). In eq. (71) the notation $\langle X\rangle_0$ means that the corresponding expectation value has been evaluated at tree level. We thus obtain the following conditions.

- At tree level

$$F_1 + F_2 + F_3 \equiv P_0 = \frac{3(N^2-2)(N^2+1)}{(3N^2-2)}\,.\tag{72}$$

The complicated form of the r.h.s. is due to the subtraction of the contribution of the protected operator $\mathcal{D}$.

- At order $g^2$ and $g^4$ the OPE analysis of (66) lead to the equations

$$g^2\ :\ F_1\eta_1 + F_2\eta_2 + F_3\eta_3 \equiv P_1 = N^2+1\,,\tag{73}$$

$$g^4\ :\ F_1\eta_1^2 + F_2\eta_2^2 + F_3\eta_3^2 \equiv P_2 = N^2 + \frac{13}{3}\,.$$

- At order $g^6$ and $g^8$ and $g^{10}$ one gets

$$
\begin{aligned}
g^6 &: \quad F_1\eta_1^3 + F_2\eta_2^3 + F_3\eta_3^3 \equiv P_3 = N^2 + b_3\,, \\
g^8 &: \quad F_1\eta_1^4 + F_2\eta_2^4 + F_3\eta_3^4 \equiv P_4 = N^2 + b_4 + \frac{c_4}{N^2}\,, \\
g^{10} &: \quad F_1\eta_1^5 + F_2\eta_2^5 + F_3\eta_3^5 \equiv P_5 = N^2 + b_5 + \frac{c_5}{N^2}\,,
\end{aligned}
\qquad (74)
$$

where $b_i$ and $c_i$ are for the moment unknown coefficients.

It is convenient to eliminate the $F_i$ from the above equations in favour of the $P_L$, obtaining the system

$$
P_{L+3} - (\eta_1+\eta_2+\eta_3)P_{L+2} + (\eta_1\eta_2+\eta_1\eta_3+\eta_2\eta_3)P_{L+1} - (\eta_1\eta_2\eta_3)P_L = 0 \qquad (75)
$$

for $L=0,1,2$. Note that this equation actually holds in general (i.e. for any value of $L$). Hence the knowledge of $P_0\ldots P_5$ completely determines all $P_L$.

In order to compute the unknown coefficients $b_i$ and $c_i$ recall that for $N=2$ there is only one possible operator, $\widehat{\mathcal{M}}$. From eq. (73) its relative anomalous dimension is determined to be $\eta_{\widehat{\mathcal{M}}} = 5/3$ for $N=2$ in agreement with eq. (38). Then from the system (74) for $N=2$ we find $b_3 = 89/9$, so that $P_3$ is completely determined, and we can express $b_4$ and $b_5$ in terms of $c_4$ and $c_5$ as follows: $b_4 = 517/27 - c_4/4$, $b_5 = 2801/81 - c_5/4$.

The knowledge of $P_3$ allows us to solve the case $N=3$, where only two operators are present with anomalous dimensions $(19\pm\sqrt{61})/18$. Then we can also fix the coefficients entering $P_4$ and $P_5$ to be $c_4 = 40/9$ and $c_5 = 1720/81$.

Substituting back these numbers into the system (75) gives for generic $N\geq 4$ the equations

$$
\begin{aligned}
\eta_1+\eta_2+\eta_3 &= \frac{8}{3}\,, \\
\eta_1\eta_2+\eta_1\eta_3+\eta_2\eta_3 &= \frac{10(2N^2-1)}{9N^2}\,, \\
\eta_1\eta_2\eta_3 &= \frac{5(3N^2-2)}{27N^2}\,,
\end{aligned}
\qquad (76)
$$

whose solutions precisely yield the values of the anomalous dimension (57) previously found in Sect. 5.

## 8  Relation with the Penrose double scaling limit

In this admittedly more speculative section we would like to discuss how one might generalize our calculations in order to make contact with the double scaling limit that corresponds to type IIB superstring around a pp-wave supported by a RR 5-form flux [27]. From the supergravity perspective this corresponds to performing a Penrose limit around a null geodesic at the center of $AdS_5 \times S^5$. The resulting maximally supersymmetric geometry [6]

$$
ds^2 = -4dx^+dx^- - \mu^2(|\vec{X}|^2 + |\vec{Y}|^2)(dx^+)^2 + (|d\vec{X}|^2 + |d\vec{Y}|^2) \qquad (77)
$$

---
[6]In our conventions, the indices of $\vec{X}$ run over $1,4,7,8$ and those of $\vec{Y}$ over $2,3,5,6$.

$$F_{+2356} = F_{+1478} = \mu \quad , \quad e^{\Phi} = g_s, \tag{78}$$

with all other fields set to zero, admits an exactly solvable worldsheet description in the Green-Schwarz formalism in the light-cone gauge [28, 45], see however [46]. In the Penrose limit the $SU(2,2|4)$ super-isometry of $AdS_5 \times S^5$ undergoes an Inonü-Wigner contraction. In particular $SO(4,2) \to SO(4)_X \times U(1)_\Delta$ and $SO(6) \to SO(4)_Y \times U(1)_J$ but at the same time a Heisenberg group, $H(8)$, emerges so that the total number of generators remains equal to 30 as for $AdS_5 \times S^5$. A similar rearrangement takes place for the 32 supersymmetry charges. In addition to the standard identifications

$$g_s = \frac{g^2}{4\pi} \quad , \quad \frac{L^2}{\alpha'} = \sqrt{g^2 N} \quad , \tag{79}$$

at large $N$ and large $J$, with $J \approx \sqrt{N}$, the relevant coupling turns out to be [28]

$$\lambda' = \frac{g^2 N}{4\pi J^2}, \tag{80}$$

where $J$ is the $U(1)_J$ charge that appears in the above decomposition of $SO(6)$ [7]. The relevant null geodesic is identified by the light-cone coordinates $x^\pm = \mu^{\mp 1}(\tau \pm \psi)/2L^2$, where $\tau$ is the global time in $AdS_5$ [3], conjugate to $\Delta$, and $\psi$ is an angular coordinate in $S^5$, conjugate to $J$. Thus in the Penrose limit the light cone momentum $P^+$ is essentially proportional to $J$, i.e.

$$p^+ = \frac{\Delta + J}{2\mu L^2} \approx \frac{J}{\mu L^2}. \tag{81}$$

Operators with $\Delta = J$ and $\Delta = J + 1$ are known to be protected, as a consequence of $SU(2,2|4)$ shortening conditions of BPS type that survive the relevant Inonü-Wigner contraction. The simplest nearly protected operators, that are expected to correspond to the lowest type IIB superstring excitations $Y_n^a Y_{-n}^b |p^+\rangle$, are of the form

$$\mathcal{A}_n^{ab} = \sum_{\ell=0}^{J} q_n^\ell tr(Z^{J-\ell} Y^a Z^\ell Y^b) \tag{82}$$

where $q_n = \exp[2\pi i n/(J+1)]$ and, in our previous notation, $Z = \phi^1$ and $Y^a = \varphi^a$ for $a = 2, 3, 5, 6$ [8].

The knowledge of the free spectrum of the light cone Hamiltonian $P^-$ gives a prediction for the 'planar' contributions to the anomalous dimensions of the operators $\mathcal{A}_n^{ab}$, i.e.

$$p^- = \frac{\mu}{2}(\Delta - J) = \mu \sqrt{1 + \frac{4\pi g_s N n^2}{J^2}} + (\text{non} - \text{planar}). \tag{83}$$

---

[7]This $U(1)_J$ does not coincide with the $U(1)_R$ in the $\mathcal{N} = 1$ decomposition of $\mathcal{N} = 4$ SYM used so far.

[8]Notice that, at variance with what is done in refs. [28, 29, 30, 31], we defined the Fourier coefficients in (82) to be $q_n = \exp[2\pi i n/(J+1)]$, and not $q_n = \exp[2\pi i n/J]$, in order to have a formula yielding $J+1$, and not $J$, linearly independent operators. This modification is at the origin of a number of further useful implications.

As noticed by [28], the effective string loop counting parameter is $J^2/N$. In [29] the first 'non-planar' corrections in SYM theory have been explicitly computed and matched with string one-loop corrections.

The obvious difference between $\mathcal{A}_n^{ab}$ and the operators we have considered in the previous sections is the presence of a large number ($J \approx \sqrt{N}$) of $Z$'s that account for their large $U(1)_J$ charge. Aside from combinatorial factors that can be elegantly deduced by resorting to a gaussian matrix model [30], we believe that most of the $Z$ fields are 'spectator' to the perturbative order one can reliably work. A subtler difference pertains to the $SO(4)$ transformation properties. In the decomposition $\mathbf{20}' = (1,1)_0 + (3,3)_0 + (2,2)_{+1} + (2,2)_{-1} + (1,1)_{+2} + (1,1)_{-2}$, the operators $\mathcal{O}_\ell^{11}$, we have mostly concentrated our attention on, transform in the $(1,1)_{+2}$. On the other hand $\mathcal{A}_n^{ab}$ belong to $(3,3)_J + (3,1)_J + (1,3)_J + (1,1)_J$. The $SU(4)$ R-symmetry of the theory should help disposing of this problem.

For generic $J$, $\mathcal{A}_n^{ab}$ can belong only to the representations which appear in the decomposition

$$[0, J, 0] \otimes [0, 1, 0] \otimes [0, 1, 0] = [0, J+2, 0] + [2, J-2, 2] + 2[1, J, 1] + \\ [2, J-1, 0] + [0, J-1, 2] + 2[1, J-2, 1] + 3[0, J, 0] + [0, J-2, 0]. \tag{84}$$

Note that only the last representation in the r.h.s. does not saturate any unitarity bound and necessarily belongs to a long multiplet. The first five, if super-primary, are protected (1/2, 1/4, 1/4, 1/8 and 1/8 BPS, respectively). Only super-descendants in these representations can acquire anomalous dimensions. The sixth and seventh representations saturate the bound $\Delta \geq 2 + k + l + m$ that leads to a linear type shortening condition of the type (23).

Although a detailed resolution of the mixing of these operators goes beyond the scope of the present investigation, we would like to argue that our computations are the building blocks for the study of their mixing properties. The $\mathcal{N} = 1$ formalism makes more transparent many laborious cancellations found to take place in [29, 30, 31]. For instance, the operator $\mathcal{A}_0^{ab}$, being totally symmetric in the $SU(4)$ indices, is a CPO belonging to the representation $[0, J+2, 0]$, hence it is protected (1/2 BPS) with $\Delta = J+2$. Moreover the symmetric, $\mathcal{A}_n^{(ab)}$, and the antisymmetric, $\mathcal{A}_n^{[ab]}$, parts of the operators defined in eq. (82) are mutually orthogonal at tree-level for all values of $n$, $N$ and $J$.

Actually one can go one step further and include also multiple trace operators. A preliminary analysis of the symmetric case for $J = 4$, including all possible mixings, shows the existence, for any $N$, of eight protected and three unprotected operators. The latter have exactly the anomalous dimensions reported in eq. (57) and turn out to be super-descendants of the operators, $\mathcal{O}_i$, identified in Sect. 5. Even though multiple trace operators decouple from single trace ones at large $N$, mixing effects may compete with 'non-planar' $(J^2/N)^2$ corrections that are dual to string loop corrections to masses. Only after disentangling operator mixing, can the comparison between gauge theory and string theory results be sensible. We will come back to this and related issues in a forthcoming publication.

Despite the success of the proposal of ref. [28], the way holography is realized in the pp-wave background is still a matter of debate [47]. This prevents a naive application of the procedure that for (asymptotically) AdS spaces has lead to 'holographic renormalization' [48]. However conformal flatness of the background, that is made manifest by the coordinate transformation [49]

$$u = \tan(x^+), \quad v = x^- - \frac{1}{2}(|\vec{X}|^2 + |\vec{Y}|^2)\tan(x^+),  \tag{85}$$

$$\vec{X}' = \frac{\vec{X}}{\cos(x^+)}, \quad \vec{Y}' = \frac{\vec{Y}}{\cos(x^+)},$$

guarantees the absence of higher derivative corrections to amplitudes with fewer than four insertions, much in the same way as in $AdS_5 \times S^5$ [50], and makes one hope that a viable string description along the lines of [46] could be not far from reach. It would then be interesting to study non-perturbative effects induced by instanton-like D-brane solutions [51] in the double scaling limit. In particular, it should be possible to extend our proof of the absence of instanton contributions to the anomalous dimensions of operators dual to string excitations. We plan to come back to these and related issues in the near future.

## 9 Conclusions

The centerpiece of this article is the resolution of the mixing of scalar operators of naive dimension 4 in the **20**$'$ of $SU(4)$ at order $g^2$ in perturbation theory. The problem is investigated in two different ways. First, by a direct orthogonalization of all possible single and double trace structures. Second, via a OPE analysis of a four point correlator computed up to order $g^4$. The agreement of the two approaches was for us an important consistency check, because the expressions found for the mixing matrix and anomalous dimensions are not rational in $N$. This certainly comes as an unexpected feature, but it is not in conflict with any fundamental property of the field theory.

Of the three unprotected operators not belonging to the Konishi multiplet that we have identified, there is one whose anomalous dimension tends to $\gamma_1^\mathcal{K}$ in the large $N$ limit. In this limit the operator under consideration can be identified with the product of the lowest Konishi operator and the lowest component of the supercurrent multiplet. We find it noteworthy that its anomalous dimension is the sum of the anomalous dimensions of its two factors, thus perhaps hinting at a deeper physical significance of singleton multiplication as a way of generating generic UIR's of $SU(2,2|4)$.

## Acknowledgements

We are grateful to Matthias Blau, Hugh Osborn, Augusto Sagnotti, Kostas Skenderis and Marika Taylor for useful conversations. M.B. thanks Soo-Jong Rey and Dan Freedman for clarifying e-mail exchanges. B.E. acknowledges many useful discussions with Emery Sokatchev and Christian Schubert. This work was supported in part by I.N.F.N., by the EC contract HPRN-CT-2000-00122, by the EC contract HPRN-CT-2000-00148, by the INTAS contract 99-0-590 and by the MURST-COFIN contract 2001-025492.

# Appendix

## The explicit calculation of the correlation function (66)

In this appendix we sketch the calculation of the connected part of the amplitude

$$A(x_1, x_2, x_3, x_4) = \langle \mathcal{C}^{11}(x_1)\mathcal{C}^\dagger_{11}(x_2) : tr(\phi^\dagger_2(x_3)\phi^2(x_3)) :: tr(\phi^\dagger_3(x_4)\phi^3(x_4)) : \rangle \qquad (86)$$

at order $g^4$. There are no connected diagrams with two vector lines, since $A$ vanishes at tree level. A direct calculation shows also that the diagrams with one chiral and one vector line sum up to zero. Hence we shall consider only diagrams with all lines corresponding to chiral propagators. We project on the lowest components of the supermultiplets, so graphs with a cubic chiral vertex with all three lines attached to external points do not contribute. Moreover, each internal chiral line gives rise to a delta function. The relevant order $g^4$ diagrams are shown in Fig. 2.

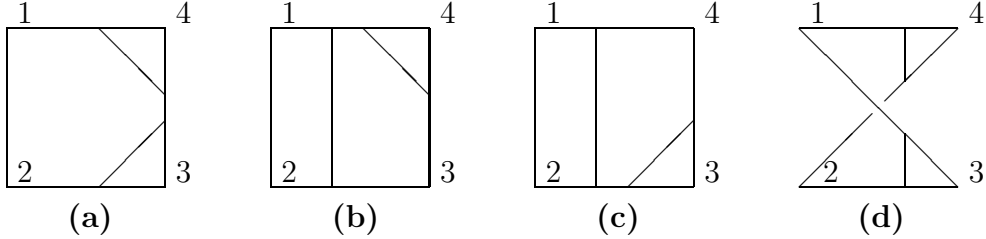

Figure 2: Feynman diagrams contributing to eq. (86)

The topologies **(a)**, **(b)** and **(c)** also occur with points 3 and 4 flipped. All diagrams have the same combinatorial weight and group factor. In order to evaluate the integral coming from diagram **(a)**, we send the argument of one of the dimension two operators, say $\mathcal{C}^\dagger_{11}$, to infinity as in eq. (65). Diagram **(d)** reduces to the product of two box integrals, so we shall concentrate on the graphs **(a)** and the sum of **(b)** and **(c)**.

## Point splitting regularization

We shall first sketch the calculation of the most complicated diagram **(a)** regularized by point-splitting as described in Sect. 2. The basic object is the massless scalar box integral defined by

$$\int \frac{d^4 x_5}{x_{15}^2 x_{25}^2 x_{35}^2 x_{45}^2} = \frac{\pi^2}{x_{13}^2 x_{24}^2} B(r, s) \; . \qquad (87)$$

A more explicit expression of $B$ is given by

$$B(r, s) = \frac{1}{\sqrt{p}} \left\{ \ln(r)\ln(s) - \left[\ln\left(\frac{r + s - 1 - \sqrt{p}}{2}\right)\right]^2 \right.$$
$$\left. - 2\text{Li}_2\left(\frac{2}{1 + r - s + \sqrt{p}}\right) - 2\text{Li}_2\left(\frac{2}{1 - r + s + \sqrt{p}}\right) \right\} , \qquad (88)$$

where

$$p = 1 + r^2 + s^2 - 2r - 2s - 2rs \qquad (89)$$

is a function of the two conformally invariant ratios

$$r = \frac{x_{12}^2 x_{34}^2}{x_{13}^2 x_{24}^2} \quad , \quad s = \frac{x_{14}^2 x_{23}^2}{x_{13}^2 x_{24}^2} \tag{90}$$

and the dilogarithm function is defined as $\text{Li}_2(z) = \sum_{n=1}^{\infty} \frac{z^n}{n^2}$.

The small $r$ limit of $B(r,s)$ is

$$\lim_{r \to 0} B(r,s) = -\ln(r) + 2 + O(r) . \tag{91}$$

The partial derivatives of $B(r,s)$ have the following form

$$\partial_r B(r,s) = -\frac{r-s-1}{p} B(r,s) - \frac{r+s-1}{pr} \ln(s) + \frac{2}{p} \ln(r) , \tag{92}$$

$$\partial_s B(r,s) = -\frac{s-r-1}{p} B(r,s) - \frac{r+s-1}{ps} \ln(r) + \frac{2}{p} \ln(s) .$$

If we introduce the notation

$$(x_{4\pm 5})^2 = (x_4 \pm \frac{\epsilon_4}{2} - x_5)^2 , \tag{93}$$

$$(x_{3\pm 6})^2 = (x_3 \pm \frac{\epsilon_3}{2} - x_6)^2 ,$$

the integral corresponding to diagram **(a)** has the form

$$I_a = \int \frac{d^4 x_6 d^4 x_5}{(x_{3+6})^2 (x_{3-6})^2 x_{56}^2 x_{15}^2 (x_{4+5})^2 (x_{4-5})^2} . \tag{94}$$

In order to evaluate it we shall add and subtract from $I_a$ the auxiliary factorized integral

$$I_1 = x_{34}^2 \int \frac{d^4 x_6}{x_{46}^2 (x_{3+6})^2 (x_{3-6})^2} \int \frac{d^4 x_5}{x_{15}^2 (x_{4+5})^2 (x_{4-5})^2 x_{35}^2} , \tag{95}$$

which in the limit of small $\epsilon_3$ and $\epsilon_4$ can be computed by the use of eqs. (87)-(91) to give

$$\lim_{\epsilon_{3,4} \to 0} I_1 = \frac{\pi^4}{x_{14}^2 x_{34}^2} \left[ -\ln\left(\frac{\epsilon_3^2}{x_{34}^2}\right) + 2 \right] \left[ -\ln\left(\frac{\epsilon_4^2 x_{13}^2}{x_{14}^2 x_{34}^2}\right) + 2 \right] . \tag{96}$$

Note that in this limit $I_1$ has logarithmic divergences, which are cancelled by the factors $(\epsilon_{3,4}^2)^{-\frac{1}{2}\gamma^{\mathcal{K}}(g^2)}$ in the definition of the renormalized Konishi operator, eq. (13). We can evaluate the integral over $x_6$ in the difference $I_a - I_1$ and take the limit $\epsilon_4 \to 0$, using again eqs. (87)-(91), obtaining

$$\lim_{\epsilon_4 \to 0} (I_a - I_1) = -\frac{\pi^2}{x_{14}^2} \mathcal{J} , \tag{97}$$

$$\mathcal{J} = \int \frac{d^4 x_6}{x_{46}^2 (x_{3+6})^2 (x_{3-6})^2} \ln\left(\frac{x_{16}^2 x_{34}^2}{x_{46}^2 x_{13}^2}\right) . \tag{98}$$

In order to calculate $\mathcal{J}$ let us first prove that it is finite in the limit $\epsilon_3 \to 0$. To this end we compute the derivative of $\mathcal{J}$ with respect to $x_1$. Since under derivation the logarithm disappears, we can express the derivative of $\mathcal{J}$ in terms of the standard box integral and take the limit $\epsilon_3 \to 0$ with the finite result

$$\lim_{\epsilon_3 \to 0} \partial_{x_1^\mu} \mathcal{J} = 2 x_{13\mu} \frac{\pi^2}{x_{13}^2 x_{34}^2} \ln\left(\frac{x_{13}^2}{x_{14}^2}\right) - \partial_{x_3^\mu}\left[\frac{\pi^2}{x_{34}^2} B\left(\frac{x_{13}^2}{x_{34}^2}, \frac{x_{14}^2}{x_{34}^2}\right)\right] . \tag{99}$$

Hence, if present, the divergent (for $\epsilon_3 \to 0$) part of $\mathcal{J}$ will be independent of $x_1$. But in the limit $x_1 \to x_4$, $\mathcal{J}$ is equal to zero, so it has a finite $\epsilon_3 \to 0$ limit. Scale invariance then allows us to write it as

$$\mathcal{J} = \frac{\pi^2}{x_{34}^2} f\left(\frac{x_{13}^2}{x_{34}^2}, \frac{x_{14}^2}{x_{34}^2}\right) . \tag{100}$$

Formula (99) gives rise to two equations for the two partial derivatives of the function $f(a, b)$ defined above. These, together with the initial condition $f(1, 0) = 0$, determine completely the function $f$, yielding

$$f(a, b) = -\frac{1}{2}\left[(a - b - 1) B(a, b) + \ln(a) \ln\left(\frac{b}{a}\right)\right] . \tag{101}$$

Inserting this solution in eqs. (100), (98) and taking into account eq. (96), we obtain the complete expression of $I_a$ (eq. (94)).

The computation of the other three diagrams **(b)**, **(c)** and **(d)** is simpler and makes use of eqs. (87)-(91) only. The final result of the calculation is given in eq. (66).

## Dimensional regularization

The calculation is done in the Euclidean regime and with the space-time dimension changed to $d = 4 + 2\epsilon$. In order to preserve the validity of equation $\nabla^2 G_0 = -\delta^{(d)}$, one has to appropriately modify the scalar propagator:

$$G_0 = \frac{1}{4\pi^2 x^2} \to \frac{\Gamma(\frac{d}{2})}{(d-2) 2\pi^{\frac{d}{2}}}\left(\frac{1}{x^2}\right)^{(\frac{d}{2}-1)} = \frac{1}{4\pi^2} \frac{\Gamma(1+\epsilon)}{\pi^\epsilon}\left(\frac{1}{x^2}\right)^{(1+\epsilon)} \tag{102}$$

The regulator is introduced in the spirit of dimensional reduction [52], *i.e.* the $x$ dependence of the superpropagators is changed as indicated above, but their $\theta$-dependence is not. One can perform the graph calculation formally since the regulator preserves the $\theta$ algebra by definition and is compatible with partial integration.

Renormalization introduces a renormalization factor $Z(g^2)$ for the Konishi operator in the operator sums : $tr(\phi_2^\dagger(x)\phi^2(x))$ : $= \mathcal{V}_2^2 + Z(g^2)\mathcal{K}_1/3$ and : $tr(\phi_3^\dagger(x)\phi^3(x))$ := $\mathcal{V}_3^3 + Z(g^2)\mathcal{K}_1/3$, with $Z|_{g^0} = 1$. The order $g^2$ part of the $Z$ factor can be found by reexpressing (14) in dimensional regularization. The order $g^4$ piece of $Z$ does not play any role in the current calculation because there are no connected tree diagrams between $\mathcal{C}^{11}, \mathcal{C}_{11}^\dagger, \mathcal{K}_\mathbf{1}$ and : $tr(\phi_I^\dagger(x)\phi^I(x))$ : for $I = 2, 3$. However, in order to obtain a finite expression for the correlator (86) we have to take into account the subtractions at order $g^4$, coming from its expression at tree level and order $g^2$.

**Diagram (a)**

The most complicated calculation concerns the contribution of this diagram. We need to compute the $x$-space integral

$$J_a(1,3,4) = \int \frac{d^4 x_5}{x_{15}^2 \, x_{45}^4} \int \frac{d^4 x_6}{x_{56}^2 x_{36}^4} \, . \tag{103}$$

We shall first evaluate the second subintegral, e.g. by the standard Feynman parameter trick. We find

$$\int \frac{d^{(4+2\epsilon)} x_6}{(x_{56}^2 \, x_{36}^4)^{(1+\epsilon)}} = \frac{\rho}{\epsilon} \left(\frac{1}{x_{35}^2}\right)^{(1+2\epsilon)} , \tag{104}$$

where

$$\rho = \frac{-\pi^{(2+\epsilon)}}{(1+2\epsilon)\Gamma(1+\epsilon)} \, . \tag{105}$$

Note that the power of the propagator in (104) is not the standard $1+\epsilon$. We split $(x_{35}^2)^{-(1+2\epsilon)} = (x_{35}^2)^{-(1+\epsilon)}(x_{35}^2)^{(-\epsilon)}$ and expand the last factor, thus obtaining three contributions to $J_a$

$$\begin{aligned} J_a(1,3,4) &= \frac{\rho}{\epsilon} \int \frac{d^4 x_5}{(x_{15}^2 \, x_{45}^4 \, x_{35}^2)^{(1+\epsilon)}} - \rho \int \frac{d^4 x_5 \, \ln(x_{35}^2)}{(x_{15}^2 \, x_{45}^4 \, x_{35}^2)^{(1+\epsilon)}} + \\ &\quad + \frac{\rho \epsilon}{2} \int \frac{d^4 x_5 \, \ln^2(x_{35}^2)}{(x_{15}^2 \, x_{45}^4 \, x_{35}^2)^{(1+\epsilon)}} + \ldots \\ &\equiv J_{a1} + J_{a2} + J_{a3} + \ldots \, . \end{aligned} \tag{106}$$

In all three pieces the remaining integral diverges and yields another simple pole in $\epsilon$. Since we need to know $J_a$ up to $O(\epsilon)$, we have to compute the first of these integrals at $O(1/\epsilon), O(1)$ and $O(\epsilon)$, the second at $O(1/\epsilon), O(1)$, but only the pole part of the third is needed and all higher terms can be ignored.

One can compute the integrals $J_{ai}$ by the method of Gegenbauer polynomials (see [53] and references therein). This approach is particularly suited to three point integrals in which only one propagator has an unusual exponent. The other two propagators are expanded in orthogonal polynomials and the angular and radial integrations are carried out. This produces two infinite series, which, however, can be written as an expansion in orthogonal polynomials of a single propagator at the expense of introducing a parametric integral [54].

We first consider the contribution $J_{a1}$. Let $x = x_{14}$, $z = x_{34}$, then for $|x| < |z|$:

$$\begin{aligned} J_{a1} \equiv \frac{\rho}{\epsilon} I_{a1} &= \frac{\rho}{\epsilon} \int \frac{d^{(4+2\epsilon)} y}{(y^4 \, (x-y)^2 \, (z-y)^2)^{(1+\epsilon)}} \\ &= \frac{\rho^2}{2\,\epsilon} \int_0^1 dt [\,(\frac{1}{z^2})^{(1+2\epsilon)} t^{(1+3\epsilon)} - (\frac{1}{x^2})^{(1+2\epsilon)} t^{-(1+\epsilon)}\,]\, \frac{1}{((z-xt)^2)^{(1+\epsilon)}} \end{aligned} \tag{107}$$

The last term in the denominator can be factorized as

$$((z-xt)^2)^{(1+\epsilon)} = (x^2)^{(1+\epsilon)} ((t-t_-)(t-t_+))^{(1+\epsilon)} , \tag{108}$$

where $t_\pm = (x \cdot z \pm \sqrt{(x \cdot z)^2 - x^2 z^2})/x^2$ [54]. The first parameter integral in (107) is regular, while the second may be performed after subtracting the singularity. The result reads

$$J_{a1} = \frac{\rho^2}{\epsilon^2} \frac{(x_{13}^2)^\epsilon}{(x_{14}^2 x_{34}^2)^{(1+2\epsilon)}} + O(1). \tag{109}$$

The finite part of the integral contains the box integral $B$ and some $\ln^2$ terms.

Remarkably, the knowledge of $J_{a1}$ is only necessary in as much as its infinities are concerned. The renormalization of the Konishi operator produces a term, $Z|_{g^2} I_{a1}$, equal to the product of $Z|_{g^2}$ and the order $g^2$ contribution to $A$, which exactly cancels the contribution $J_{a1}$ [9].

Let us now focus on $J_{a2}$. We shall write the integral as

$$J_{a2} = -\rho \int \frac{d^{(4+2\epsilon)} x_5 \, \ln(x_{35}^2)}{(x_{15}^2 \, x_{35}^2 \, x_{45}^4)^{(1+\epsilon)}} = \frac{\rho}{1+\epsilon} \partial_\beta \int \frac{d^{(4+2\epsilon)} x_5}{(x_{15}^2 \, (x_{35}^2)^\beta \, x_{45}^4)^{(1+\epsilon)}} \bigg|_{\beta=1}. \tag{110}$$

Define

$$K_{a2} = \int \frac{d^{(4+2\epsilon)} x_5}{(x_{15}^2)^{(1+\epsilon)} (x_{45}^2)^{(1+2\epsilon)} (x_{35}^2)^{\beta(1+\epsilon)}}. \tag{111}$$

We can isolate the divergence in $J_{a2}$ by writing

$$\left(\frac{x_{14} \partial_4}{1 + 2\epsilon} - 1\right) K_{a2} = (x_{14} \partial_4 - 1) \int \frac{d^4 x_5}{x_{15}^2 \, x_{45}^2 \, (x_{35}^2)^\beta} + O(\epsilon), \tag{112}$$

since the integral is finite if $\beta$ is close to one. Swapping integration and differentiation in the l.h.s. we derive (the integrals are finite as long as the regulator is not removed)

$$\left(\frac{x_{14} \partial_4}{1 + 2\epsilon} - 1\right) K_{a2} \tag{113}$$

$$= x_{14}^2 \int \frac{d^{(4+2\epsilon)} x_5}{(x_{15}^2 \, x_{45}^4)^{(1+\epsilon)} (x_{35}^2)^{\beta(1+\epsilon)}} - \int \frac{d^{(4+2\epsilon)} x_5}{(x_{15}^2)^\epsilon (x_{45}^4)^{(1+\epsilon)} (x_{35}^2)^{\beta(1+\epsilon)}}.$$

Let us call the second term in the last line $L_{a2}$. Expanding $(x_{15}^2)^{-\epsilon}$ up to first order in $\epsilon$ and recombining the pole parts of the resulting two integrals one finds

$$L_{a2} = -\frac{1}{(x_{14}^2)^\epsilon} \int \frac{d^{(4+2\epsilon)} x_5}{(x_{45}^4)^{(1+\epsilon)} (x_{35}^2)^{\beta(1+\epsilon)}} + O(\epsilon). \tag{114}$$

We substitute this expression into (113) and equate with (112). After dividing by $x_{14}^2$, we obtain

$$\int \frac{d^{(4+2\epsilon)} x_5}{(x_{15}^2 \, (x_{35}^2)^\beta \, x_{45}^4)^{(1+\epsilon)}} = \frac{1}{(x_{14}^2)^{(1+\epsilon)}} \int \frac{d^{(4+2\epsilon)} x_5}{((x_{35}^2)^\beta \, x_{45}^4)^{(1+\epsilon)}} \tag{115}$$

$$+ \frac{x_{14} \partial_4 - 1}{x_{14}^2} \int \frac{d^4 x_5}{x_{15}^2 \, (x_{35}^2)^\beta \, x_{45}^2} + O(\epsilon). \tag{116}$$

---

[9]The finite part of the $Z$ factor leads to a term of the type $b \, I_{a1}$, which also produces a first order pole in $\epsilon$. By a judicious choice of $b$ this term is cancelled by $J_{a1}$.

The first term in the r.h.s. contains a divergence, but as a two-point structure it is readily calculable. We compute the second term in the r.h.s. again by means of Gegenbauer polynomials. As above, we introduce a one parameter integral to rewrite the result of the angular integration in closed form. One finds for $\beta = 1 + \Delta$

$$\int \frac{d^4 x_5}{x_{15}^2 \, x_{45}^2 \, (x_{35}^2)^{(1+\Delta)}} = \left(1 - \frac{\Delta}{2}(\ln(x_{13}^2) + \ln(x_{34}^2))\right) \int \frac{d^4 x_5}{x_{15}^2 \, x_{35}^2 \, x_{45}^2} + O(\Delta^2), \tag{117}$$

which is sufficient to calculate the first order parametric derivative in (110). Finally, we employ the identity

$$\frac{x_{14}\partial_4 - 1}{x_{14}^2} \int \frac{d^4 x_5}{x_{15}^2 \, x_{35}^2 \, x_{45}^2} = \frac{\pi^2}{x_{14}^2 \, x_{34}^2} \ln\left(\frac{x_{14}^2}{x_{13}^2}\right). \tag{118}$$

Collecting terms we find

$$\begin{aligned}
J_{a2} = & -\frac{\rho^2}{\epsilon} \frac{\ln(x_{34}^2)}{(x_{14}^2)^{(1+\epsilon)}(x_{34}^2)^{(1+2\epsilon)}} + \\
& \frac{\pi^4}{2} \frac{1}{x_{14}^2 \, x_{34}^2} \bigg[\left(1 - \frac{x_{14}^2}{x_{13}^2} - \frac{x_{34}^2}{x_{13}^2}\right) B\left(\frac{x_{14}^2}{x_{13}^2}, \frac{x_{34}^2}{x_{13}^2}\right) + \\
& (\ln(x_{13}^2) + \ln(x_{34}^2)) \ln\left(\frac{x_{14}^2}{x_{13}^2}\right)\bigg] + O(\epsilon).
\end{aligned} \tag{119}$$

As for $J_{a3}$, since we learned above how to extract a simple pole, without any further calculation we get

$$J_{a3} = \frac{\pi^4}{2} \frac{\ln(x_{34}^2)^2}{x_{14}^2 \, x_{34}^2} + O(\epsilon). \tag{120}$$

**Diagrams (b) and (c)**

It is convenient to compute directly the sum of these graphs, since compensations between the two terms lead to a simpler result. The $x$-space integrals are of the form

$$J_b + J_c = \int \frac{d^4 x_5}{x_{15}^2 \, x_{35}^2} \int \frac{d^4 x_6}{x_{56}^2 x_{36}^2 x_{46}^4} + (3 \leftrightarrow 4). \tag{121}$$

The second subintegral is divergent. It is given by $I_{a1} = \epsilon J_{a1}/\rho$ which was calculated already in eqs. (107)-(109). Its coordinate behaviour at orders $O(1/\epsilon)$ and $O(1)$ is essentially $1/(x_{34}^2 x_{45}^2)$, which does not lead to overlapping divergences in the second integration. We can then expand $I_{a1}$ obtaining (one can neglect terms of order $O(\epsilon)$ and higher)

$$\begin{aligned}
J_b + J_c & = \frac{\rho}{\epsilon} \frac{1}{(x_{34}^2)^{(1+2\epsilon)}} \int \frac{d^{(4+2\epsilon)} x_5}{(x_{15}^2 \, x_{35}^2 \, x_{45}^2)^{(1+\epsilon)}} \left[\left(\frac{x_{45}^2}{x_{35}^2}\right)^\epsilon + \left(\frac{x_{35}^2}{x_{45}^2}\right)^\epsilon\right] \\
& = \frac{\rho}{\epsilon} \frac{2}{(x_{34}^2)^{(1+\epsilon)}} \int \frac{d^{(4+2\epsilon)} x_5}{(x_{15}^2 \, x_{35}^2 \, x_{45}^2)^{(1+\epsilon)}} + 2\pi^2 \frac{\ln(x_{34}^2)}{x_{34}^2} \int \frac{d^4 x_5}{x_{15}^2 \, x_{35}^2 \, x_{45}^2}
\end{aligned} \tag{122}$$

Note that the first term contains trilogarithms but is again identically cancelled by $Z|_{g^2}$ times an order $g^2$ diagram. The second integral is once again the box $B$. The end result of the calculation is in exact agreement with (66).

# References

[1] O. Aharony, S.S. Gubser, J. Maldacena, H. Ooguri and Y. Oz, "Large N field theories, string theory and gravity", *Phys. Rept.* **323** (2000) 183, `hep-th/9905111`.

[2] D.Z. Freedman and P. Henry-Labordere, "Field theory insight from the AdS/CFT correspondence", `hep-th/0011086`.

[3] E. D'Hoker and D.Z. Freedman, "Supersymmetric Gauge Theories and the AdS/CFT Correspondence", `hep-th/0201253`.

[4] M. Bianchi, "(Non-)perturbative tests of the AdS/CFT correspondence", *Nucl. Phys.* **B** *Proc. Suppl.* **102** (2001) 56, `hep-th/0103112`.

[5] W. Skiba, "Correlators of short multi-trace operators in $\mathcal{N}=4$ supersymmetric Yang-Mills", *Phys. Rev.* **D60** (1999) 105038, `hep-th/9907088`.

[6] S. Lee, S. Minwalla, M. Rangamani and N. Seiberg, "Three-Point functions of chiral operators in $D=4, \mathcal{N}=4$ SYM at Large $N$", *Adv. Theor. Math. Phys.* **2** (1998) 697, `hep-th/9806074`; E. D'Hoker, D.Z. Freedman and W. Skiba, "Field Theory Tests for Correlators in the AdS/CFT Correspondence", *Phys. Rev.* **D59** (1999) 045008, `hep-th/9807098`; B. Eden, P.S. Howe, P.C. West, "Nilpotent invariants in $\mathcal{N}=4$ SYM", *Phys. Lett.* **B463** (1999) 19, `hep-th/9905085`

[7] H. Liu and A.A. Tseytlin, "Dilaton-fixed scalar correlators and $AdS_5 \times S^5$ - SYM correspondence", *JHEP* **9910** (1999) 003, `hep-th/9906151`; E. D'Hoker, D.Z. Freedman, S.D. Mathur, A. Matusis and L. Rastelli, "Extremal Correlators in the AdS/CFT Correspondence", `hep-th/9908160`; M. Bianchi and S. Kovacs, "Non-renormalisation of extremal correlators in $\mathcal{N}=4$ SYM theory", *Phys. Lett.* **B468** (1999) 102, `hep-th/9910016`; B. Eden, P.S. Howe, C. Schubert, E. Sokatchev and P.C. West, "Extremal correlators in four-dimensional SCFT", *Phys. Lett.* **B472** (2000) 323, `hep-th/9910150`.

[8] J. Erdmenger and M. Perez-Victoria, "Non-renormalization of next-to-extremal correlators in $\mathcal{N}=4$ SYM and the AdS/CFT correspondence", *Phys. Rev.* **D62** (2000) 045008, `hep-th/9912250`;

[9] M. Bianchi, M.B. Green, S. Kovacs and G.C. Rossi, "Instantons in supersymmetric Yang-Mills and D-instantons in IIB superstring theory", *JHEP* **9808** (1998) 013, `hep-th/9807033`; N. Dorey, T.J. Hollowood, V.V. Khoze, M.P. Mattis and S. Vandoren, "Multi-instanton calculus and the AdS/CFT correspondence in $\mathcal{N}=4$ superconformal field theory", *Nucl. Phys.* **B552** (1999) 88, `hep-th/9901128`.

[10] K. Intriligator, "Bonus symmetries of $\mathcal{N}=4$ super-Yang-Mills correlation functions via AdS Duality", *Nucl. Phys.* **B551** (1999) 575, `hep-th/9811047`; K. Intriligator and W. Skiba, "Bonus symmetry and the Operator Product Expansion of $\mathcal{N}=4$ super-Yang-Mills", *Nucl. Phys.* **B559** (1999) 165, `hep-th/9905020`.

[11] J. Maldacena, "Wilson loops in large N field theories", *Phys. Rev. Lett.* **80** (1998) 4859, `hep-th/9803002`; S. Rey and J. Yee, "Macroscopic strings as heavy quarks in large N gauge theory and anti-de Sitter supergravity", *Eur. Phys. J.* **C22** (2001) 379, `hep-th/9803001`.

[12] N. Drukker, D.J. Gross and H. Ooguri, "Wilson loops and minimal surfaces", *Phys. Rev.* **D60** (1999) 125006, `hep-th/9904191`; H. Ooguri, "Wilson loops in large N theories", *Class. Quant. Grav.* **17** (2000) 1225, `hep-th/9909040`; N. Drukker and D.J. Gross, "An exact prediction of $\mathcal{N} = 4$ SUSYM theory for string theory", *J. Math. Phys.* **42** (2001) 2896, `hep-th/0010274`.

[13] J.K. Erickson, G.W. Semenoff and K. Zarembo, "Wilson loops in $\mathcal{N} = 4$ supersymmetric Yang-Mills theory", *Nucl. Phys.* **B582** (2000) 155, `hep-th/0003055`; G.W. Semenoff and K. Zarembo, "More exact predictions of SUSYM for string theory", *Nucl. Phys.* **B616** (2001) 34, `hep-th/0106015`.

[14] J. Plefka and M. Staudacher, "Two loops to two loops in $\mathcal{N} = 4$ supersymmetric Yang-Mills theory", *JHEP* **0109**, (2001) 031, `hep-th/0108182`; G. Arutyunov, J. Plefka and M. Staudacher, "Limiting Geometries of Two Circular Maldacena-Wilson Loop Operators", *JHEP* **0112** (2001) 014, `hep-th/0111290`.

[15] M. Bianchi, M.B. Green and S. Kovacs, "Instantons and BPS Wilson loops" `hep-th/0107028`, M. Bianchi, M.B. Green and S. Kovacs, "Instanton corrections to circular Wilson loops in $\mathcal{N} = 4$ supersymmetric Yang-Mills", *JHEP* **0204** (2002) 040, `hep-th/0202003`.

[16] B. Eden, A.C. Petkou, C. Schubert and E. Sokatchev, "Partial non renormalisation of the stress-tensor four-point function in $\mathcal{N} = 4$ SYM and AdS/CFT", *Nucl. Phys.* **B607** (2001) 191, `hep-th/0009106`; E. D'Hoker, J. Erdmenger, D.Z. Freedman and M. Perez-Victoria, "Near-extremal correlators and vanishing supergravity couplings in AdS/CFT", *Nucl. Phys.* **B589** (2000) 3, `hep-th/0003218`.

[17] F.A. Dolan and H. Osborn, "Implications of $\mathcal{N} = 1$ superconformal symmetry for chiral fields", *Nucl. Phys.* **B593** (2001) 599, `hep-th/0006098`; "Conformal four point functions and the Operator Product Expansion", *Nucl. Phys.* **B599** (2001) 459, `hep-th/0011040`; "Superconformal symmetry, correlation functions and the Operator Product Expansion", *Nucl. Phys.* **B629** (2002) 3, `hep-th/0112251`.

[18] M. Bianchi, S. Kovacs, G.C. Rossi and Ya.S. Stanev, "Properties of the Konishi multiplet in $\mathcal{N} = 4$ SYM theory", JHEP **0105** (2001) 042, `hep-th/0104016`.

[19] B. Eden, P.S. Howe, C. Schubert, E. Sokatchev and P.C. West, "Four point functions in $\mathcal{N} = 4$ supersymmetric Yang-Mills theory at two loops", *Nucl. Phys.* **B557** (1999) 355, `hep-th/9811172`; B. Eden, P.S. Howe, C. Schubert, E. Sokatchev and P.C. West, "Simplifications of four point functions in $\mathcal{N} = 4$ supersymmetric Yang-Mills theory at two loops", *Phys. Lett.* **B466** (1999) 20, `hep-th/9906051`; B. Eden, P.S. Howe, A. Pickering, E. Sokatchev and P.C. West, "Four point functions in $\mathcal{N} = 2$

superconformal field theories", *Nucl. Phys.* **B581** (2000) 523, `hep-th/0001138`; B. Eden, C. Schubert and E. Sokatchev, "Three-Loop Four-Point Correlator in $\mathcal{N}=4$ SYM", *Phys. Lett.* **B482** (2000) 309, `hep-th/0003096`.

[20] M. Bianchi, S. Kovacs, G.C. Rossi and Ya.S. Stanev, "On the logarithmic behaviour in $\mathcal{N}=4$ SYM theory", *JHEP* **9908** (1999) 020, `hep-th/9906188`.

[21] M. Bianchi, S. Kovacs, G.C. Rossi and Ya.S. Stanev, "Anomalous dimensions in $\mathcal{N}=4$ SYM theory at order $g^4$", *Nucl. Phys.* **B584** (2000) 216, `hep-th/0003203`.

[22] G. Arutyunov, B. Eden, A.C. Petkou and E. Sokatchev, "Exceptional non-renormalization properties and OPE analysis of chiral four-point functions in $\mathcal{N}=4$ SYM$_4$", *Nucl. Phys.* **B620** (2002) 380, `hep-th/0103230`; G. Arutyunov, B. Eden and E. Sokatchev, "On non-renormalization and OPE in superconformal field theories", *Nucl. Phys.* **B619** (2001) 359, `hep-th/0105254`; B. Eden and E. Sokatchev, "On the OPE of 1/2 BPS short operators in $\mathcal{N}=4$ SCFT(4)", *Nucl. Phys.* **B618** (2001) 259, `hep-th/0106249`; S. Ferrara and E. Sokatchev, "Universal properties of superconformal OPEs for 1/2 BPS operators in $3 \leq D \leq 6$", *New Jour. Phys.* **4** (2002) 2, `hep-th/0110174`; P.J. Heslop and P.S. Howe, "OPEs and three-point correlators of protected operators in $\mathcal{N}=4$ SYM", *Nucl. Phys.* **B626** (2002) 265, `hep-th/0107212`.

[23] P.J. Heslop and P.S. Howe, "On Harmonic Superspaces and Superconformal Fields in Four Dimensions", *Class. Quant. Grav.* **17** (2000) 3743, `hep-th/0005135`; "Harmonic superspaces and superconformal fields", `hep-th/0009217`; P.J. Heslop, "Superfield representations of superconformal groups", *Class. Quant. Grav.* **19** (2002) 303, `hep-th/0108235`.

[24] S. Ferrara and A. Zaffaroni, "$\mathcal{N}=1,2$ 4D Superconformal Field Theories and Supergravity in $AdS_5$", *Phys. Lett.* **B431** (1998) 49, `hep-th/9803060`; "Bulk Gauge Fields in AdS Supergravity and Supersingletons", `hep-th/9807090`; "Superconformal Field Theories, Multiplet Shortening and the $AdS_5 \times S^5$ Correspondence", `hep-th/9908163`.

[25] P.J. Heslop and P.S. Howe, "A note on composite operators in $\mathcal{N}=4$ SYM", *Phys. Lett.* **B516** (2001) 367, `hep-th/0106238`

[26] G. Arutyunov, S. Frolov and A.C. Petkou, "Perturbative and instanton corrections to the OPE of CPO's in $\mathcal{N}=4$ SYM$_4$", *Nucl. Phys.* **B602** (2001) 238, Erratum *ibid* **B609** (2001) 540, `hep-th/0010137`; G. Arutyunov, S. Frolov and A.C. Petkou, "Operator product expansion of the lowest weight CPO's in $\mathcal{N}=4$ SYM(4) at strong coupling", *Nucl. Phys.* **B586** (2000) 547, `hep-th/0005182`.

[27] M. Blau, J. Figueroa-O'Farrill, C. Hull and G. Papadopoulos, "A new maximally supersymmetric background of IIB superstring theory", *JHEP* **0201** (2002) 047, `hep-th/0110242`; M. Blau, J. Figueroa-O'Farrill, C. Hull and G. Papadopoulos, "Penrose limits and maximal supersymmetry", *Class. Quant. Grav.* **19** (2002) L87,

`hep-th/0201081`; M. Blau, J. Figueroa-O'Farrill and G. Papadopoulos, "Penrose limits, supergravity and brane dynamics", `hep-th/0202111`.

[28] D. Berenstein, J.M. Maldacena and H. Nastase, "Strings in flat space and pp-waves from $\mathcal{N}=4$ super Yang Mills", *JHEP* **0204** (2002) 013, `hep-th/0202021`.

[29] N.R. Constable, D.Z. Freedman, M. Headrick, S. Minwalla, L. Motl, A. Postnikov and W. Skiba, "pp-wave string interactions from perturbative Yang-Mills theory", `hep-th/0205089`.

[30] C. Kristjansen, J. Plefka, G.W. Semenoff and M. Staudacher, "A new double-scaling limit of $\mathcal{N}=4$ super Yang-Mills theory and pp-wave strings", `hep-th/0205033`.

[31] D.J. Gross, A. Mikhailov and R. Roiban, "Operators with large R charge in $\mathcal{N}=4$ Yang-Mills theory", `hep-th/0205066`.

[32] L. Brink, J. Scherk and J.H. Schwarz, "Supersymmetric Yang-Mills theories", *Nucl. Phys.* **B121** (1977) 77; F. Gliozzi, D.I. Olive and J Scherk, "Supersymmetry, supergravity and the dual spinor model", *Nucl. Phys.* **B122** (1977) 253.

[33] M. Grisaru, M. Roček and W. Siegel, "Zero value of the three-loop $\beta$ function in $\mathcal{N}=4$ supersymmetric Yang-Mills theory", *Phys. Rev. Lett.* **45** (1980) 1063; W.E. Caswell and D. Zanon, "Zero three-loop beta-function in $\mathcal{N}=4$ supersymmetric Yang-Mills theory", *Nucl. Phys.* **B182** (1981) 125.

[34] S. Penati, A. Santambrogio and D. Zanon, "Two-point functions of chiral operators in $\mathcal{N}=4$ SYM at order $g^4$", *JHEP* **9912** (1999) 006, `hep-th/9910197`.

[35] S. Kovacs, "A perturbative re-analysis of $\mathcal{N}=4$ supersymmetric Yang–Mills theory", `hep-th/9902047`; "$\mathcal{N}=4$ supersymmetric Yang-Mills theory and the AdS/SCFT correspondence", Ph.D. Thesis, `hep-th/9908171`.

[36] L. Andrianopoli and S. Ferrara, "K-K excitations on $AdS_5 \times S^5$ as $\mathcal{N}=4$ primary superfields", *Phys. Lett.* **B430** (1998) 248, `hep-th/9803171`; "Non-chiral primary superfields in the $\mathrm{AdS}_{d+1}/\mathrm{CFT}_d$ correspondence", *Lett. Math. Phys.* **46** (1998) 265, `hep-th/9807150`; "On short and long $SU(2,2|4)$ multiplets in the AdS/CFT correspondence", *Lett. Math. Phys.* **48** (1999) 145, `hep-th/9812067`.

[37] D. Anselmi, D.Z. Freedman, M.T. Grisaru and A.A. Johansen, "Universality of the Operator Product Expansions of $\mathrm{SCFT}_4$", *Phys. Lett.* **B394** (1997) 329, `hep-th/9608125`; "Non-perturbative formulas for central functions of supersymmetric gauge theories", *Nucl. Phys.* **B526** (1998) 543, `hep-th/9708042`; D. Anselmi, J. Erlich, D.Z. Freedman and A.A. Johansen, "Positivity Constraints on Anomalies in Supersymmetric Gauge Theories", *Phys. Rev.* **D57** (1998) 7570, `hep-th/9711035`; D. Anselmi, "The $\mathcal{N}=4$ quantum conformal algebra", *Nucl. Phys.* **B541** (1999) 369, `hep-th/9809192`.

[38] V.K. Dobrev and V.B. Petkova, "All positive energy unitary irreducible representations of extended conformal supersymmetry", *Phys. Lett.* **B162** (1985) 127.

[39] M. Flato and C. Fronsdal, "One massless particle equals two Dirac singletons: Elementary particles in a curved space. 6", *Lett. Math. Phys* **2** (1978) 421; "On dis and racs", *Phys. Lett.* **B97** (1980) 236; "Quantum field theory of singletons: The rac", *J. Math. Phys.* **22** (1981) 1100; "Quarks or singletons?", *Phys. Lett.* **B172** (1986) 412.

[40] M. Gunaydin and N. Marcus, "The spectrum of the $S^5$ compactification of the chiral $\mathcal{N} = 2, D = 10$ supergravity and the unitary supermultiplets of $U(2, 2/4)$", *Class. Quant. Grav.* **2** (1985) L11.

[41] L. Andrianopoli, S. Ferrara, E. Sokatchev and B. Zupnik, "Shortening of primary operators in $\mathcal{N}$-extended SCFT$_4$ and harmonic superspace analyticity", *Adv. Theor. Math. Phys.* **3** (1999) 1149, hep-th/9912007; S. Ferrara and E. Sokatchev, "Short representations of $SU(2, 2|\mathcal{N})$ and harmonic superspace analyticity", *Lett. Math. Phys.* **52** (2000) 247, hep-th/9912168; "Superconformal interpretation of BPS states in AdS geometries", *Int. J. Theor. Phys.* **40** (2001) 935, hep-th/0005151.

[42] A.S. Galperin, E.A. Ivanov, V.I. Ogievetsky and E.S. Sokatchev, "Harmonic superspace", *University. Press* (Cambridge – UK, 2001) page 306.

[43] M.F. Sohnius, "Bianchi identities for supersymmetric gauge theories", *Nucl. Phys.* **B136** (1978) 461; "Supersymmetry and central charges", *Nucl. Phys.* **B138** (1978) 109; W. Siegel, "On-shell O(N) supergravity in superspace", *Nucl. Phys.* **B177** (1981) 325; P.S. Howe, K.S. Stelle and P.K. Townsend, *Nucl. Phys.* **B191** (1981) 445; *Nucl. Phys.* **B192** (1981) 332.

[44] K. Konishi, "Anomalous supersymmetry transformation of some composite operators in SQCD", *Phys. Lett.* **B135** (1984) 195; K. Konishi and K. Shizuya, "Functional integral approach to chiral anomalies in supersymmetric gauge theories", *Nuovo Cim.* **A90** (1985) 111.

[45] R.R. Metsaev, "Type IIB Green-Schwarz superstring in plane wave Ramond-Ramond background", *Nucl. Phys.* **B625** (2002) 70, hep-th/0112044; R.R. Metsaev and A.A. Tseytlin, "Exactly solvable model of superstring in plane wave Ramond-Ramond background", hep-th/0202109.

[46] N. Berkovits, "Conformal field theory for the superstring in a Ramond-Ramond plane wave background", *JHEP* **0204** (2002) 037, hep-th/0203248; "Covariant quantization of the superstring", *Int. J. Mod. Phys.* **A16** (2001) 801, hep-th/0008145.

[47] S.R. Das, C. Gomez and S.J. Rey, "Penrose limit, spontaneous symmetry breaking and holography in pp-wave background", hep-th/0203164; E. Kiritsis and B. Pioline, "Strings in homogeneous gravitational waves and null holography", hep-th/0204004; R.G. Leigh, K. Okuyama and M. Rozali, "pp-waves and holography", hep-th/0204026.

[48] See *e.g.* , M. Bianchi, D.Z. Freedman and K. Skenderis, "How to go with an RG flow", *JHEP* **0108** (2001) 041, `hep-th/0105276`; "Holographic renormalization", `hep-th/0112119` and references therein.

[49] D. Berenstein and H. Nastase, "On lightcone string field theory from super Yang-Mills and holography", `hep-th/0205048`.

[50] T. Banks and M.B. Green, "Non-perturbative effects in $AdS_5 \times S^5$ string theory and $d = 4$ SUSY Yang-Mills", *JHEP* **9805** (1998) 002, `hep-th/9804170`.

[51] M. Billo and I. Pesando, "Boundary states for GS superstrings in an Hpp-wave background", `hep-th/0203028`; K. Skenderis and M. Taylor, "Branes in AdS and pp-wave spacetimes", `hep-th/0204054`; O. Bergman, M.R. Gaberdiel and M.B. Green, "D-brane interactions in type IIB plane-wave background", `hep-th/0205183`.

[52] W. Siegel, "Supersymmetric dimensional regularization via dimensional reduction", *Phys. Lett.* **B84** (1979) 193.

[53] K.G. Chetyrkin, A.L. Kataev and F.V. Tkachov, "New approach to evaluation of multiloop Feynman integrals: The Gegenbauer $x$-space technique", *Nucl. Phys.* **B174** (1980) 345.

[54] D.Z. Freedman, K. Johnson and J.I. Latorre, "Differential regularization and renormalisation: A new method of calculation in quantum field theory", *Nucl. Phys.* **B371** (1992) 353